\documentclass[%
reprint,superscriptaddress,amsmath,amssymb,aps,prb,longbibliography
]{revtex4-2}

\usepackage{graphicx}
\usepackage{dcolumn}
\usepackage{bm}
\usepackage{amssymb}
\usepackage{amsmath}
\usepackage{float}
\usepackage{subfigure}
\usepackage{xcolor}
\usepackage{color, soul}
\usepackage{tabularx}
\usepackage{array}
\renewcommand\arraystretch{1.25}
\usepackage[flushleft]{threeparttable}
\usepackage{amsthm}
\usepackage{amsfonts}
\usepackage{braket}
\usepackage{natbib, hyperref}
\hypersetup{colorlinks=true,linkcolor=blue, filecolor=blue,urlcolor=blue,citecolor=blue,}
\usepackage{enumitem}

\usepackage [english]{babel}
\usepackage [autostyle, english = american]{csquotes}
\MakeOuterQuote{"}

\usepackage[version=4]{mhchem}
\usepackage{chemmacros}
\chemsetup{
    modules = all,
    formula = mhchem
}

\usepackage{chemformula}

\usepackage{soul}

\begin{document}

\pagestyle{plain}

\title{An Atomistically Informed Device Engineering (AIDE) Method Realized: \\ A case study in GaAs}

\author{Leopoldo Diaz}
\email{Lndiaz@sandia.gov}
\noaffiliation
\author{Harold P. Hjalmarson}\noaffiliation%
\author{Jesse J. Lutz}\noaffiliation
\author{Peter A. Schultz}\noaffiliation
 
\affiliation{%
 Sandia National Laboratories, Albuquerque, New Mexico 87185, USA
}%

\begin{abstract}
Radiation-induced defects can have a significant impact on the longevity and performance of semiconductor devices. We present an Atomistically Informed Device Engineering (AIDE) method that integrates first-principles defect properties and experimentally measured parameters into a device model to dynamically simulate the defect chemistry in semiconductors. For a silicon-doped gallium arsenide (GaAs) material, we showcase three capabilities: (i) Fermi level $E_F$ movement including its component electron and hole Fermi levels, (ii) dynamical charge equilibration with the arsenic vacancy serving as an example, and a (iii) diffusion-driven reaction between Coulomb attracted gallium interstitial ($Ga_i$) and arsenic vacancy ($v_{As}$). Governed by charge carrier reactions, the electron and hole Fermi levels remained dissimilar until equilibrium was achieved at $E_F\approx1.32$ eV. The equilibrium Fermi level was verified by successfully identifying $v_{As}^{3-}$ as the most populated charge state within the arsenic vacancy defect. Lastly, a Coulomb attraction, created by the shifted Fermi level and the charge equilibration process, between $Ga_i^{1+}$ and $v_{As}^{3-}$ resulted in the formation of a doubly negative gallium antisite ($Ga_{As}^{2-}$). The AIDE method can access experimentally inaccessible short-time and low-concentration regimes, is generalizable to other more complex systems (e.g., indium gallium arsenide), and, after solving open problems in GaAs, will serve as a virtual experiment to bound estimates for difficult-to-measure physical quantities.
\end{abstract}

\maketitle
\section{Introduction}
Materials and devices in space environments encounter a variety of energetic particles including protons, electrons, and ions that result in the creation of atomic defects--displacement damage, i.e., defects created by non-ionizing energy loss (NIEL). These radiation-induced defects (vacancies and interstitials) are detrimental to performance and accumulate to cause system/device failure. Understanding defect evolution is crucial for assessing radiation sensitivities, and is, especially, important for maximizing device longevity for remote applications such as in satellite electronics. 

Indium gallium arsenide (InGaAs), an advantageous photodiode for satellite communications systems in the infrared range, has attracted the attention of researchers for approximately 20 years \cite{walters1992,shaw1993,shaw1993_2,hugon1995,barde2000,gilard2018,nuns2020}, with several authors attributing degradation to displacement damage \cite{barde2000,gilard2018,nuns2020}. However, despite the availability of deep-level transient spectroscopy (DLTS), defect characterization in InGaAs remains challenging and has inspired several recent works \cite{gilard2018,nuns2020}. Compounded by the technical difficulties associated with obtaining useful measurements in this relatively narrow gap material, additional challenges stem from the inability of experimental techniques to probe the rapid and low-temperature annealing (defect reactions) that occurs in InGaAs \cite{nelson2020}. 
Atomistic defect simulations are essential for identifying and characterizing radiation-induced defects, but are inadequate for simulating annealing behavior. In this paper, we propose a defect-informed multiscale approach to modeling the dynamical defect evolution in this and other semiconductor materials. 

Unfortunately, a lack of reliable experimental and theoretical works in InGaAs makes this ternary semiconductor an impractical material for model development. However, many experimental \cite{bourgoin1988,pons_bourgoin1985,Stievenard1990} and theoretical works~\cite{baraff1985,zhang1991,schultz2009,schultz2015,schultz2022} exist for chemically similar, and simpler, GaAs. Similarly, the defect physics in GaAs mimics many of the same issues, including questions about radiation-induced defect identification \cite{taghizadeh2018} and annealing. While the two materials share many similar open questions concerning radiation response, GaAs has a greater availability of quality experimental and theoretical studies, including a complete and accurate Rosetta Stone of point defect levels calculated from first-principles density functional theory (DFT) \cite{schultz2009,schultz2015,schultz2016,schultz2022}. This makes GaAs an ideal material to develop and exercise our multiscale dynamical model.

Despite decades of interest, many important phenomena involving radiation-induced defects and defect evolution in GaAs remain unexplained, including:
\begin{itemize}
    \item A physical mechanism explaining why the Ga vacancy is “invisible” to experimental probes, despite DFT predicting it to be stable and immobile~\cite{schultz2009,El-Mellouhi2006};
    \item The defect responsible for the recently discovered E3b peak observed by high resolution Laplace DLTS \cite{taghizadeh2018};
    \item The cause of the Fermi level pinning \cite{Walukiewicz1988} in GaAs, which some hypothesized to be due to oxygen-related defects \cite{colleoni2013,colleoni2016,luo2024}.
\end{itemize}

Answers to these defect-related questions continue to elude theoretical and experimental methods because they lack the capability to investigate the dynamical defect evolution that is initiated by an irradiation event. Although Ab initio Molecular Dynamics (AIMD) and quantum chemistry techniques are capable of performing defect reactions, reaching sufficient accuracy and meaningful timescales is computationally infeasible with modern hardware and algorithms. 

Unraveling this dynamical response requires a deeper understanding of the chemical evolution of defects, including the simultaneous occurrence of multiple reaction types between charged species over meaningful timescales.

In this work, we present an Atomistically Informed Device Engineering (AIDE) method that uses reliable DFT and experimental defect properties within the Radiation Effects in Oxides and Semiconductors (REOS) continuum device code in a multiscale approach to model the dynamical defect evolution (Fig.~\ref{fig:fig1}). In Section \ref{sec:irradiation}, we provide an overview of the general electron irradiation process, emphasizing the key steps relevant to this work. A description of the AIDE model is given in Section \ref{sec:aide}, followed by a brief mathematical description of the REOS code with commentary on the relevant quantities in Section \ref{sec:calc}, and an explanation of parameter selection in Section \ref{sec:params}. In Section \ref{sec:results}, we demonstrate the capabilities of our AIDE method by simulating the repercussions of electron irradiation on a Si-doped GaAs sample including (i) the creation, movement, and convergence of the quasi-Fermi levels; (ii) the charge-state concentration of each defect which verifies the converged Fermi level location in the bandgap; and (iii) a Coulomb-driven defect-defect reaction (diffusion-driven) between a mobile interstitial and an immobile vacancy. Finally, we conclude with summarizing remarks, future works, and aspirations for the AIDE method. 

\begin{figure*}[t]
    \centering
\includegraphics[width=0.95\textwidth]{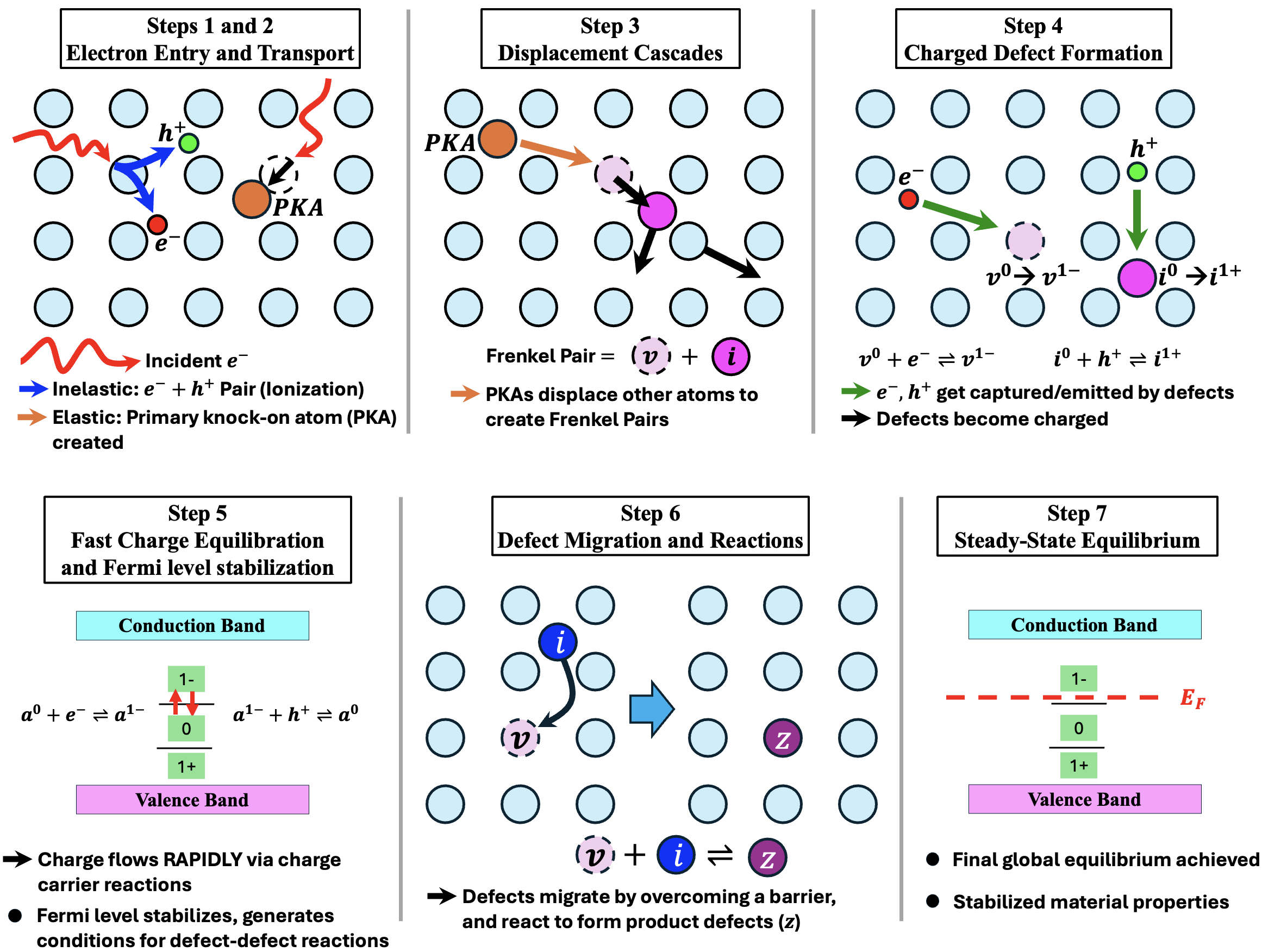}
    \caption{Illustrative representation of the electron irradiation process and the sequential steps involved. The representation starts with the electron entering the material, causing the generation of defects, and ends when the system reaches a new steady-state equilibrium.}
    \label{fig:iradmech}
\end{figure*}

\section{Irradiation Process}
\label{sec:irradiation}
Irradiation is the process by which materials/devices are exposed to high-energy particles or electromagnetic radiation, resulting in interactions that can alter their physical, chemical, structural, and/or electronic properties. High-energy particles initiate complex sequences of non-linear dynamical processes that evolve simultaneously in time. The irradiation process is inherently multiscale and multiphysics, with interactions spanning orders of magnitude in both time and space. The complexity of modeling this process is further compounded by factors such as the type of irradiation, level of fluence, and material properties. Focusing on high-energy electron irradiation, we provide a concise summary of the sequential steps that occur during an irradiation event (see Fig. \ref{fig:iradmech}).

\begin{enumerate}[label=Step \arabic*:, align=left, leftmargin=*, labelwidth=!, widest=Step 10:]
    \item \textit{Electron Entry} \newline High energy electrons approach the surface and penetrate the material through a "window". 
    \item \textit{Electron Transport} \newline Once inside the material, the electron follows a zigzag (stochastic) path due to multiple simultaneous interactions, including inelastic and elastic collisions, that result in the formation of electron and hole quasi-Fermi levels.
    \begin{enumerate}[label=(\alph*)]
        \item \ul{Inelastic Collisions}: These collisions occur when incident electrons interact with host atom electrons resulting in a loss of kinetic energy, the creation of electron-hole pairs (ionization), and an initial charge imbalance. This charge imbalance is temporary, localized, and caused by the generation of free charge carriers. The imbalance is transient and resolves through processes like recombination or carrier transport--the material remains globally charge neutral.  
        \item \ul{Elastic Collisions}: These collisions involve the incident electron scattering off host lattice atoms, transferring momentum and energy (total kinetic energy is conserved). If the electron has sufficient energy, the host atom can be knocked out of its lattice position, creating a primary knock-on atom (PKA). The PKA can then initiate a displacement cascade. 
    \end{enumerate}
    \item \textit{Displacement Cascades} \newline
    As the electron scatters within the material, more PKAs are generated and displacement cascades are initiated with:
    \begin{enumerate}[label=(\alph*)]
        \item \ul{Low-Energy PKAs} Resulting in isolated point defects--including Frenkel pairs (interstitial-vacancy pairs)--and secondary knock-on atoms which can then displace additional atoms. 
        \item \ul{High-Energy PKAs} Leading to the formation of defect clusters--large aggregations of displaced atoms.
    \end{enumerate}
    \item \textit{Charged Defect Formation} \newline 
    Isolated point defects generated by low-energy PKAs interact with charge carriers through defect-carrier reactions. These reactions include defects capturing and emitting electrons and holes, leading to the formation of charged defects, which contribute to charge imbalance.
    
    \item \textit{Initial Fast Charge Equilibration} \newline
    Once defects have become charged, the system redistributes charge carriers to restore overall neutrality--charge equilibration. This \textit{rapid} charge equilibration includes defects continuing to capture and emit charge carriers, driving the system toward equilibrium. During this process, the electron and hole quasi-Fermi levels will fluctuate until equilibrium is achieved. When the quasi-Fermi levels converge, equilibrium is achieved and the most populated charge states for each defect is established. This initial stabilization generates the conditions for Coulomb-driven defect-defect reactions, i.e., oppositely charged defects react with one another producing product defects.
    
    \item \textit{Defect Migration and Reactions} \newline 
    As charge equilibration continues, defect migration becomes more prominent. Defects react with one another through diffusion-driven defect-defect reactions. Defects that are Coulombically attracted to one another react much faster. We emphasize that defect migration and reactions occur slower than charge equilibration processes but occur simultaneously as they influence each other. 
    \item \textit{Steady-State Equilibrium} \newline Eventually, the system reaches a new steady state--final global equilibrium--where defect migration and defect-defect reactions have stabilized, allowing the material properties to achieve stability. 

 \end{enumerate}
To date, no single, unified model can simulate all these steps. Instead, distinct methods are used to model different steps. Density functional theory (DFT) is highly effective for simulating defect properties, including their charge states (Step 4), transition levels (Step 5), and migration barriers (Step 6). However, DFT is limited in its ability to dynamically simulate these processes at meaningful time scale and at necessary length scale. In the following section, we present an approach that uses reliable DFT data and experimental data to dynamically simulate Steps 4-7--defect evolution--in a material/device. 

\begin{figure*}[t!]
    \centering
    \includegraphics[width=.95\textwidth]{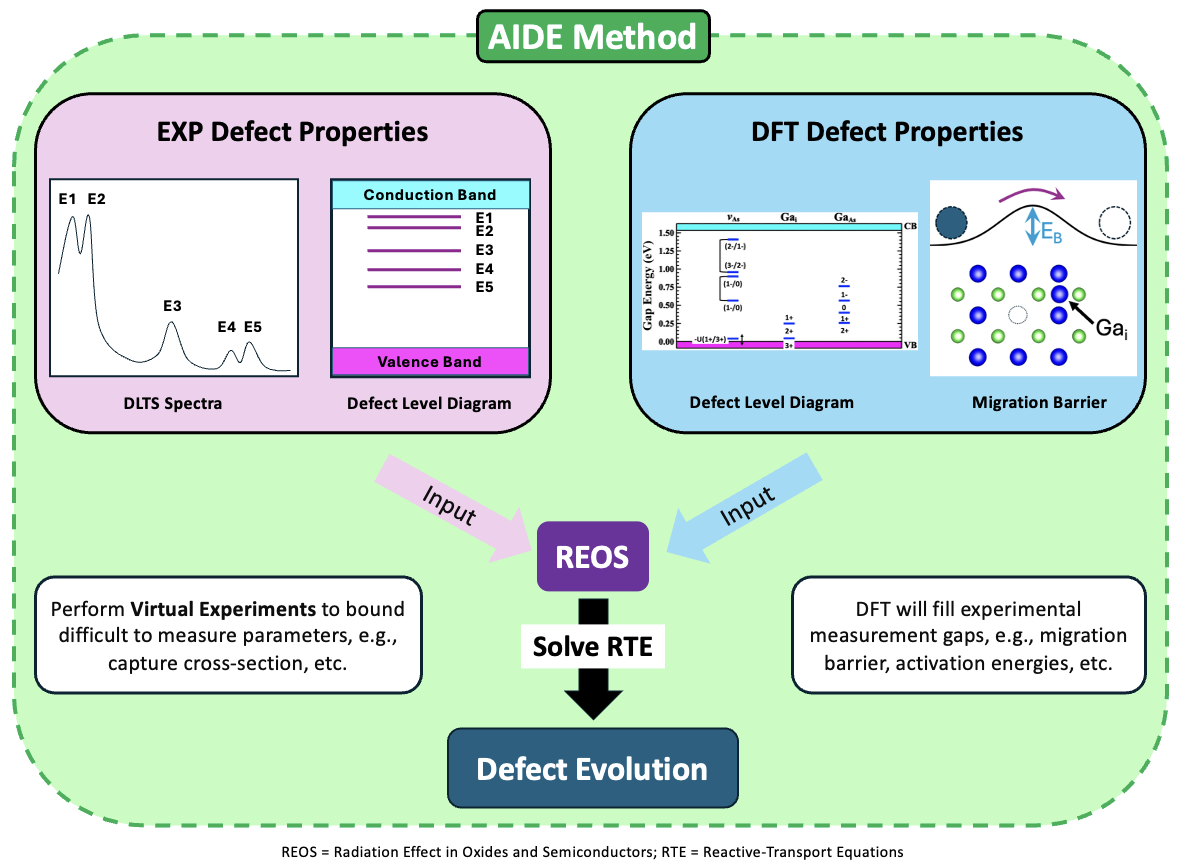}
    \caption{Illustrative representation of the Atomically Informed Device Engineering (AIDE) method and its inclusion in the REOS device software suite. The experimental (EXP) defect properties includes illustrations of a DLTS spectra and a mapping of the peaks onto the GaAs bandgap. The DFT defect properties include a defect level diagram extracted from \cite{schultz2009} and a illustrative representation of the migration barrier.}
    \label{fig:fig1}
\end{figure*}

\section{AIDE Method Development}
\label{sec:aide}
Generally, semiconductor device theory codes require several input parameters such as capture cross-sections, diffusion properties, etc. Deep-level transient spectroscopy (DLTS) has proven invaluable in characterizing defects, providing critical information such as the number of defects present, their activation energies, their relative concentration, and their energy level location within the band gap \cite{pons_bourgoin1985,taghizadeh2018,lang1974}. Ideally, these parameters would be measured using some experimental technique and used as input in REOS. However, several parameters are difficult to measure and extremely variable depending on the method, conditions, or sample. For instance, defects detected by DLTS and low-frequency noise spectroscopy (LFNS) often do not agree, with one technique detecting a greater or lesser number of defects than the other, and the reported defect energy levels often differ substantially. Such discrepancies have been observed even in well-studied silicon \cite{luo2024}. The causes for these discrepancies remain unknown. These inconsistencies introduce additional uncertainty and gaps in device modeling efforts. Fortunately, first-principles DFT can reliably predict many critical defect properties--including atomic structure, charge transitions (defect levels), and migration energy barriers (diffusion activation energies) [Fig.~\ref{fig:fig1}]--thereby filling many of these parameter gaps left by experimental limitations. 

The AIDE method uses defect properties from experiment and atomistics (DFT) and uses them as input properties in the device code REOS. Using this information augmented by physical intuition where needed, REOS solves the reactive-transport equations (Section \ref{sec:calc}) and predicts the dynamical evolution of all species, including electrons, holes, defects, impurities, and dopants (Fig. \ref{fig:fig1}) with a continuum scale simulation. The use of experimental and DFT properties ensures a physically accurate treatment of the dynamical evolution of defects in a material. 

This AIDE method aims to simulate the dynamical evolution of defects by employing physically reasonable parameters in a device code. This approach is computationally efficient and provides valuable insight into experimentally inaccessible regimes, making it a viable alternative to computationally prohibitive brute-force AIMD and time-dependent quantum chemistry methods. Note, two other independently-developed (also from Sandia National Laboratories), yet similar, defect-informed multiscale simulation approaches have been published \cite{myers2008,wampler2015,charon}; however, these methods were geared toward understanding defect clusters--neutron damage. This work focuses on using the AIDE method for isolated point defects created by electron irradiation.

\section{Calculation Details}
\label{sec:calc}
Developed over several decades by Hjalmarson \cite{loubriel1997,kambour2000,meyers2025,hjalmarson2003,pease2008,hjalmarson2008,tierney2018}, REOS is a continuum research code whose primary goal is to develop radiation damage models for material/device simulations. The REOS code has previously enabled studies of the following:
\begin{itemize}
    \item Collective impact ionization of lock-on in gallium arsenide photoconductive switches (PCSS) \cite{loubriel1997, kambour2000}, which can qualitatively predict switch performance. Recently, this model has been extended to wide-bandgap gallium nitride switches \cite{meyers2025}; 
    \item The technologically important problem of enhanced low dose-rate sensitivity (ELDRS) in bipolar transistors. The model successfully predicted experimentally observed dose-rate dependence on interface traps \cite{hjalmarson2003, pease2008, hjalmarson2008};
    \item The physical mechanism of leakage current in metal-insulator-metal (MIM) systems which included simulating electron transport mechanisms (e.g., band-to-band tunneling and band-to-defect tunneling) through the metal-oxide interface \cite{tierney2018}.
\end{itemize}
 These theoretical models developed with REOS have shown at least qualitative agreement with experimental observations. The current work extends the radiation physics capabilities in REOS to perform atomistically-informed continuum-scale calculations of reactive transport to simulate dynamical defect evolution in irradiated materials. This dynamic capability enables the study of many simultaneous reactions over experimentally inaccessible short-time and low-defect concentration regimes.

REOS solves reaction transport calculations \cite{shockley1949,roosbroeck1950} for a specimen (e.g., slab, diode, or other semiconducting device) in one or two dimensions. The specimen is inserted into an electrical circuit with an arbitrary number of contacts. The specimen can be comprised of defects, dopants, and charge carriers ($e^-$/$h^+$) with the charge carriers and imperfections undergoing transport and chemical processes.

The temporal evolution of each chemical species $i$ [electrons, holes, defects, dopants, etc.] is governed by a reactive transport equation (RTE). The chemical species participate in two chemical reaction types (i) electron/hole ($e^-/h^+$) capture and emission reactions (charge carrier reactions) and (ii) defect-defect reactions (diffusion-driven reactions). Before exploring each reaction type, a general discussion will be given on the RTE and the makeup of the REOS suite. The RTE is given by 

\begin{equation}
\label{eqn:rte}
    \frac{\partial c_i(\textbf{r}, t)}{\partial t} = - \nabla \cdot \textbf{J}_{i}(\textbf{r}, t) + \sum_j \gamma_{ij}r_j 
\end{equation}
where $c_i(\textbf{r}, t$) is the concentration of species $i$, $\textbf{J}_{i}(\textbf{r},t)$ is the species particle current density (flux of species $i$), $r_j$ is the reaction rate for reaction $j$, and $\gamma_{ij}$ is the stoichiometric coefficient of species $i$ in reaction $j$. The first term on the right side of Eqn. \ref{eqn:rte} is the transport term and the term furthest to the right is the reactive term. The species current density is

\begin{equation}
    \label{eqn:flux2}
    J_i = -q_ic_i\mu_i \nabla \Phi_i
\end{equation}
where $q_i$, $\mu_i$,  and $\Phi_i$ are the species charge, mobility, and electrochemical potential of species $i$, respectively. Using the Einstein relation, the mobility is related to the diffusion coefficient ($D_i$) as follows:
\begin{equation}
    \label{eqn:einstein}
    D_i = \frac{\mu_i kT}{q_i},
\end{equation}
where $k$ and T are the Boltzmann constant and temperature. By substituting Eqn. \ref{eqn:einstein} into Eqn. \ref{eqn:flux2}, the species current density becomes
\begin{equation}
    J_i=-\frac{q^2_ic_iD_i}{k T}\nabla \Phi_i.
\end{equation}

The electron ($J_e$) and hole ($J_h$) current densities are given by

\begin{equation}
    J_e=-\frac{q_e^2 n_e D_e}{k T}\nabla \Phi_e
\end{equation}

\begin{equation}
    J_h=-\frac{q_h^2 n_hD_h}{k T}\nabla \Phi_h
\end{equation}
where $n_e$ and $n_h$ are the electron and hole concentrations, respectively. The electrochemical potential is defined as the sum of the electrostatic potential [$\phi(\textbf{r})$] and the chemical potential [$\nu_i(\textbf{r})$]

\begin{equation}
\label{eqn:electro}
    \Phi_i(\textbf{r}) = q_i \phi(\textbf{r}) + \nu_i(\textbf{r}).
\end{equation}
Species diffusion in semiconductors occurs when there are gradients in the electrochemical potential. These gradients can arise from differences in species concentration, applied voltage, or temperature, leading to changes in the system's electrostatic potential, chemical potential, or thermodynamics. 

The electrostatic potential $\phi(\textbf{r})$,
\begin{equation}
\label{eqn:poisson}
    \nabla^2 \phi(\textbf{r}) = - \frac{\rho(\textbf{r})}{\varepsilon}
\end{equation}
is calculated from Poisson's equation (Eqn. \ref{eqn:poisson}) with the total charge density given by 
\begin{equation}
    \rho = \sum_i z_i q_i c_i
\end{equation}
which depends on the species charge number $z_i$, $q_i$, and $c_i$. The dielectric coefficient $\varepsilon$ is equal to the product of the relative permittivity $\kappa$ and the vacuum permittivity $\varepsilon_0$ ($\varepsilon = \kappa \varepsilon_0$). 

The chemical potential is a fundamental quantity in REOS simulations, and semiconductor device simulations in general, as it provides key information about the Fermi level. The chemical potential for electrons ($\nu_e$) and holes ($\nu_h$) is given by
\begin{align}
    \nu_e (\textbf{r}) = -q\phi(\textbf{r}) + E_F^e \\
    \nu_h (\textbf{r}) = q\phi(\textbf{r}) + E_F^h .
\end{align}
Immediately after an irradiation event, the system is knocked out of equilibrium. As a result, the system cannot be described by a single Fermi level. Instead, separate dynamical quasi-Fermi levels are computed to describe electrons $E_F^e$ and holes $E_F^h$. First proposed by Shockley \cite{shockley1949}, these quasi-Fermi levels provide important information about how charge flows throughout the system. The spatially dependent quasi-Fermi levels are defined by 
\begin{align}
\label{eqn:fermie}
    E_F^e = kT ln\Bigg(\frac{n_e}{N_C}\Bigg) \\ 
    \label{eqn:fermih}
    E_F^h = -kT ln\Bigg(\frac{n_h}{N_V}\Bigg) .
\end{align}
The electron and hole quasi-Fermi levels are referenced to the conduction and valence band edges, respectively.

The effective density of states in the conduction ($N_C$) and valence ($N_V$) bands is dependent on the effective masses of electrons ($m_e^*$) and holes ($m_h^*$), 


\begin{align}
    N_C =2\Bigg(\frac{m_e^* kT}{2\pi \hbar^2}\Bigg)^{3/2} \\
    N_V =2\Bigg(\frac{m_h^* kT}{2\pi \hbar^2}\Bigg)^{3/2}
\end{align}
where $\hbar$ is Planck's constant. 
Combining several of these equations (Eqn.~\ref{eqn:electro} into Eqn.~\ref{eqn:flux2}), one can transform Eqn.~\ref{eqn:flux2} into a flux equation for electrons $J_e$ and holes $J_h$
\begin{equation}
\label{eqn:dde}
    J_e = qn_e\mu_e E_{field} +qD_e \nabla n_e
\end{equation}

\begin{equation}
\label{eqn:ddh}
    J_h = qn_h\mu_h E_{field} -qD_h \nabla n_h
\end{equation}
where $E_{field}$ is an applied electric field and $D_e$ ($D_h$) is the diffusion coefficient for electrons (holes). In Eqns. \ref{eqn:dde} and \ref{eqn:ddh}, the first term on the right represents drift and the second term accounts for diffusion. For point defects, the flux equation becomes
\begin{equation}
    J_i = qc_i\mu_i E +qD_i \nabla c_i.
\end{equation}
The diffusion coefficient $D_i$ is computed using 
\begin{equation}
\label{eqn:diffusion}
    D_i = D_{0i} e^{-\frac{E_{mi}}{kT}}
\end{equation}
where $D_{0i}$ and $E_{mi}$ are the diffusion prefactor, and migration barrier, respectively, for species $i$.  Since chemical reactivity between defects is highly probable, $D_i$ and $r_j$ are critical to accurately represent the physics of semiconducting materials. In this work, there is no applied field or voltage; however, the approach can straightforwardly incorporate bias effects if desired.

\begin{figure}[t]
    \centering
    \includegraphics[width=0.9\columnwidth]{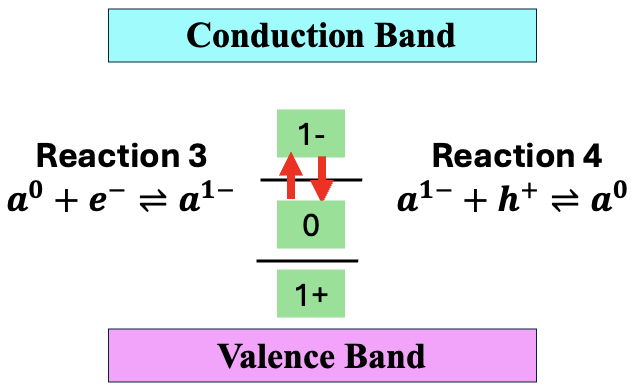}
    \caption{Qualitative defect level diagram for a generic defect $a$ in three different charge states (-1, 0, and +1) and two transition levels. Red arrows indicate how charge flows via charge carrier capture and emission reactions.}
    \label{fig:eh}
\end{figure}

\subsection{Charge Flow} 
Immediately following the primary displacement damage, charge flows through the system until an equilibrium is obtained---Fermi level stabilization. The flow of charge within the material is the fastest process, occurs before atomic diffusion and defect reactions, and is critical to determine the Fermi level location within the band gap. The REOS suite utilizes two reaction types to simulate charge flow, including electron ($e^-$) and hole ($h^+$) capture and emission reactions (charge carrier reactions). The charge carrier reactions are a fundamental reaction type in the REOS suite. Not only do these reactions enable charge flow throughout the system, but also, subsequently, maintain charge neutrality via charge-conserving reactions such as 

\begin{reactions*}
    &\llap{Reaction 1} & a^{1+} + e^- & <=>  a^0 \\
    &\llap{Reaction 2} & a^{0}  + h^+ & <=>  a^{1+} \\
    &\llap{Reaction 3} & a^{0}  + e^- & <=>  a^{1-} \\
    &\llap{Reaction 4} & a^{1-}  + h^+ & <=>  a^{0}
\end{reactions*}
where $a$ is a generic defect in three charge states ($1+$, $0$, and $1-$). An illustration of this process is shown in Fig.~\ref{fig:eh} where Reactions 3 and 4 are shown for the $0$ and $1-$ charge states of a generic defect $a$. This carrier capture process applies to all defect charge states that span the entire band gap. Depending on the number of defects and charge states per defect, the number of charge carrier reactions can become very large. Using Reactions 1 and 2 as examples, the reaction rates $k_f^{a^{1+}}$ and $k_f^{a^0}$ are given by

\begin{align}
\label{eqn:cross1}
    &k_{f}^{a^{1+}} & = v_{te}\sigma&\Big\{-[a^{1+}]n_ee^{-\frac{E_f}{kT}}+N_C[a^0]e^{-\frac{E_r}{kT}}\Big\} \\
    \label{eqn:cross2}
    &k_{f}^{a^{0}} &= v_{th}\sigma&\Big\{-[a^{0}]n_he^{-\frac{E_f}{kT}}+N_V[a^{1+}]e^{-\frac{E_r}{kT}}\Big\}
\end{align}
where $v_{te}$ ($=\sqrt{\frac{3kT}{m_e^*}}$) and $v_{th}$ ($=\sqrt{\frac{3kT}{m_h^*}}$) are the $e^-$ and $h^+$ thermal velocities, respectively and $\sigma$ is the capture cross-section. The $E_{f}$ and $E_r$ are activation energies---defect levels---and are measured by experiment and/or predicted by DFT.

\subsection{Diffusion-Driven Reactions}
As discussed in Section \ref{sec:calc}, species diffusion is driven by an electrochemical potential gradient that dictates where mobile species want to move. Atomistically, diffusion of mobile defects typically requires overcoming an activation barrier. For neutral and charged interstitial species, this involves diffusing through the lattice until the species is trapped or reacts with another defect. Once charge flows through the system, the relative populations of each defect charge state is determined by the Fermi level, with the majority of the defect population settling into charge states near the Fermi level position. In a physical system, mobile defects diffuse throughout the lattice and trigger chemical reactions via diffusion-driven reactions. If the defect species are mobile and charged, these reactions are accelerated by Coulombic attraction or supressed by Coulombic repulsion. 

In REOS, these defect-defect chemical reactions take the form

\begin{reactions*}
    &\llap{Reaction 5} & aA + bB & <=>  cC + dD
\end{reactions*}
where A, B, C, and D are generic defects (A and B are the reactants and C and D are the products) and a, b, c, and d are their coefficients, respectively. In Eqn. \ref{eqn:rte}, $r_j$ is defined by the mass-action law
\begin{equation}
\label{eqn:ddrate}
    r_j = k_{jf}[A_j]^{a_j}[B_j]^{b_j}-k_{jr}[C_j]^{c_j}[D_j]^{d_j}
\end{equation}
where $k_{jf}$ and $k_{jr}$ are the forward and reverse reaction rates, respectively, for reaction $j$. In principle, reverse reactions would and could be treated in the REOS suite. However, in practice, these reverse reactions are unlikely to result in a more stable configuration at room temperature (exothermic) and, even if considered, are unlikely to produce a substantial concentration of product. Therefore, reverse reactions were not included in this work. The forward reaction rate ($k_{jf}$) for diffusion-driven reactions is given by


\begin{equation}
\label{eqn:ddreact}
    k_{jf}= 4\pi R_{eff} [D_{A}+D_{B}]
\end{equation}
where R$_{eff}$ is an effective radius that represents the probability that two defects will interact, i.e., the smaller the radius means the less likely they will react. 

The $D_A$ and $D_B$ are the diffusion coefficients for defect A and B and are determined by Eqn. \ref{eqn:diffusion}. A diffusion coefficient of zero ($D=0$ cm/s$^2$) is chosen for immobile species.  

Several semiconducting parameters are not well characterized; however, achieving the correct order of magnitude for parameters such as the capture cross-section, suffices in developing a physically accurate representation of material systems. Methods used for parameter selection are discussed in Section \ref{sec:params}.

A complete physical and mathematical description of the REOS code will be provided in a separate work. 

\section{Virtual Experiments, Parameter Selection, and approximations}
\label{sec:params}
The AIDE method is a multiscale approach that bridges the atomistic and continuum device scales. Coupling across scales makes parameter selection challenging because some are difficult to measure, do not translate, or are experimentally and theoretically unknown, in which case physical and chemical intuition is the primary guide. For example, species concentrations, capture cross-sections, and diffusion properties are all difficult to measure or compute and have required special attention. In addition to experimental and theoretical guidance, \textit{virtual experiments} were performed to gain insight and intuition into the material behavior and parameters (see Sec. \ref{sec:virtualexp}). 

In the following subsections, a description of parameter selection is given for the difficult-to-determine parameters used in our simulations of irradiated Si-doped GaAs.






\subsection{Virtual Experiments}
\label{sec:virtualexp}
Virtual experiments are computational simulations designed to mimic physical experiments, enabling exploration and testing of parameters in a controlled, virtual environment. Using the AIDE method, virtual experiments were conducted for the following purposes:
\begin{enumerate}
    \item Determining sensitivity for physical parameters, providing insight into the physics of the problem, including how small or large changes affect the behavior of the system. 
    \item Assessing how well uncertain experiment and DFT perform with respect to multiscale modeling, offering insight into how useful atomistic tools are for providing reliable data.
    \item Understanding how well (1) and (2) compare to device experiments. This comparison informs how effectively a reduced compact device model incorporates the essential physical phenomena associated with the device response being investigated. 
\end{enumerate} 
Virtual experiments give a sense of how sensitive the results are to changes in crucial parameters and how well the parameters reflect physical reality. In future work, \textit{virtual experiments} will be used to guide experimentalists by suggesting ranges, setups, and conditions--where to look--as they seek measurements of these difficult-to-measure parameters. The use of virtual experiments to bound difficult-to-measure parameters and guide experimenters is a major goal of the AIDE method.

\subsection{Irradiation event and Defect Distribution}
\label{subsec:distr}
The distribution of defects in a real material/device depends on several factors including, to name a few, the radiation source (electron, neutron, ion, etc), fluence, and the properties of the material/device. High energy electrons ($> 10$ keV) collide elastically with atoms, displacing them from their lattice sites, creating vacancy and interstitial point defects. Following their creation, charge is redistributed to achieve an equilibrium state--charge equilibration. After this rapid process, defects migrate and can react with other defects, impurities, and/or dopants. 

In GaAs, irradiation with 1-MeV electrons does not produce the broad U- and L- bands \cite{pons_bourgoin1985}---characteristic of clustered damage---and, instead, produces point defects that are considered spatially uniform~\cite{fleming2010}. DLTS experiments have observed five electron traps (E1-E5) with E1-E3 identified as point defects \cite{schultz2009,schultz2015} and E4-E5 associated with point defect complexes based on their location in the band gap. For these reasons, a uniform defect distribution was assumed in these simulations and was sufficient to illustrate the most important radiation effects. A non-uniform distribution can and will be implemented in future works. 

\begin{figure}[h!]
    \centering
    \includegraphics[width=\columnwidth]{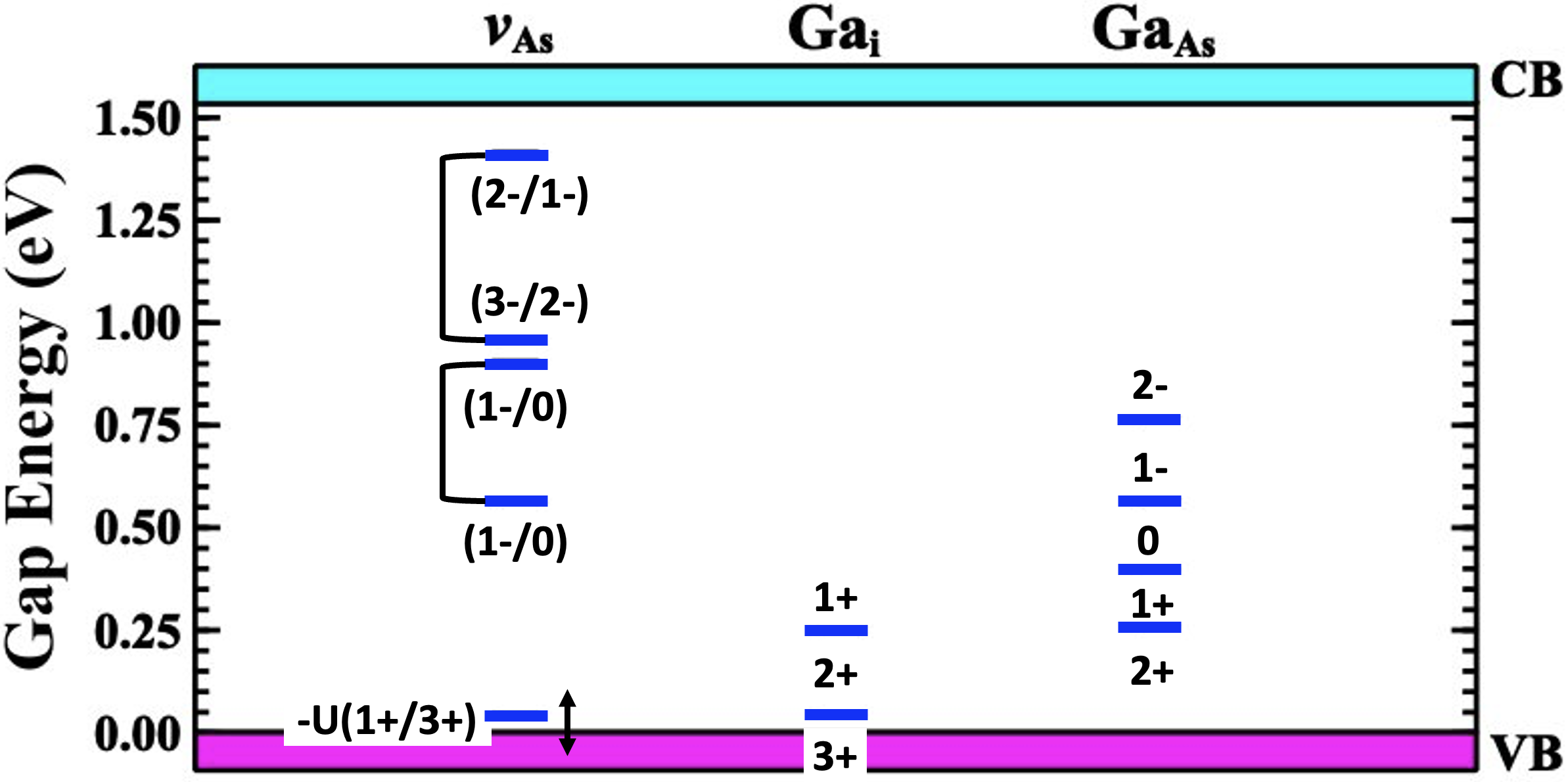}
    \caption{First-principles computed defect levels of $v_{As}$, $Ga_i$, and $Ga_{As}$ in irradiated Si-doped GaAs \cite{schultz2009,schultz2016}. The brackets indicate -U behavior.}
    \label{fig:fig2}
\end{figure}

\subsection{Charge state transition levels (Defect Levels)}
Radiation-induced defects create defect levels--transition states--in the band gap with each defect existing in numerous charge states. DLTS measurements provide key information on the number of defects in the bandgap and their location in the bandgap. In GaAs, theory has proven pivotal in the identification of experimentally observed traps, including the well-known EL2 \cite{dabrowski1988,chadi1988} in as-grown GaAs. Furthermore, the Rosetta Stone of point defects in GaAs has facilitated the characterization of radiation-induced E1-E2 as the divacancy ($v_{Ga}v_{As}$ \cite{schultz2015}) and E3 as the arsenic vacancy ($v_{As}$ \cite{schultz2009}). Defect levels for $v_{As}$, $Ga_i$, and $Ga_{As}$ are provided in Fig. \ref{fig:fig2}.

DFT simulations have provided essential defect level information for several other materials, including Si \cite{schultz2006} and GaN \cite{edwards2022}. The average absolute error for DFT-computed defect levels is approximately 0.2 eV, with maximum departure of 0.2 eV. Virtual experiments with AIDE of the dynamical response show minimal sensitivity to defect levels within 0.2 eV.  This result indicates that achievable accuracy in DFT can be sufficient to meet the requirements for accurate multiscale simulations of device response.  The use of the AIDE method in virtual experiments to assess the sensitivities in simulations parameters can quantitatively assess the needed accuracy in those parameters such as defect levels.


\subsection{Species Concentrations}
\label{sub:specCon}
An accurate experimental assessment of defect concentrations is generally not available. In electron-irradiated n-GaAs, experimentally measured defect concentrations are never comprehensive (nor definitive), but provide enough information to enable reasonable estimates. The initial conditions for the REOS calculations, the starting defect and dopant populations, are chosen to be roughly consistent with experimental measurements, and guided by the need to maintain internal consistency between the relative defect populations.

The GaAs sample is doped with Si at an experimentally relevant concentration \cite{taghizadeh2018} of $4.5\times10^{15}$ cm$^{-3}$ with two charge states $Si_{Ga}^0$ and $Si_{Ga}^{1+}$. The defect concentrations are chosen to match the general GaAs DLTS defect concentrations, which range between $10^{14}-10^{16}$~cm$^{-3}$. Using the DLTS concentration range enables the use of peak heights to relate the defect concentrations with each other. As shown in Fig. \ref{fig:fig1}, DLTS spectra \cite{pons1980} have revealed that E1-E2 ($v_{Ga}v_{As}$ \cite{schultz2015}) are taller than E3 ($v_{As}$ \cite{schultz2009}). Since peak heights are proportional to defect populations, vacancy populations are chosen such that $[v_{Ga}v_{As}]>[v_{Ga}]=[v_{As}]$ \cite{lang1974}. The single vacancies are assumed to be relatively equal. In contrast, no GaAs interstitial has been experimentally observed, and any knowledge of their population sizes remains unknown. As a result, for this proof-of-principle work, they were selected to be equal ($Ga_i=As_i$) and greater than the number of vacancies. The defect populations used in this work are provided in Table \ref{tab:defectpops}. 

\begin{table}[h]
\renewcommand{\arraystretch}{1.3}
\caption{\label{tab:defectpops} Population sizes and migration barriers for the radiation-induced defects in n-GaAs.  Doping concentration is $[Si_{Ga}]=4.5\times10^{15}$ cm$^{-3}$.}
\centering
\begin{tabularx}{\columnwidth}{lcc}
\hline
\bfseries Defect&
\bfseries Concentration ($cm^{-3}$)&
\bfseries Migration Barrier (eV)\\
\hline\hline
$As_{i}$ & $3.0\times10^{15}$ & $\approx0.5$ \cite{pons_bourgoin1985}\\
$v_{As}$ & $4.0\times10^{14}$ & Immobile\\
$Ga_{i}$& $3.0\times10^{15}$ & $\approx1.0$ \cite{schultz2009}, 1.06 \cite{pichler}\\
$v_{Ga}$ & $4.0\times10^{14}$ & Immobile\\
$v_{Ga}v_{As}$ & $1.0\times10^{15}$ & Immobile\\
\hline
\end{tabularx}
\end{table}

Small changes in the doping and defect populations do not alter the results in a meaningful way. However, large enough changes, decreasing (increasing) dopant (defect) concentrations, can cause a Fermi level shift, thereby altering the defect chemistry of the system. Further exploration is needed to understand the effect of doping on the defect chemistry, i.e., the interplay between carrier-defect reactions and defect-defect reactions during changes in the doping population.

\subsection{Capture Cross-section}
The carrier capture cross-section $\sigma$ (Eqns. \ref{eqn:cross1} and \ref{eqn:cross2}) is a physical parameter that describes the likelihood that an electron or hole (charge carrier) will be captured by a defect, dopant, and/or impurity. It is quantified as an effective area [$\sigma$ (cm$^2$)] around a defect that will interact with charge carriers. However, measuring and computing the capture cross-section is difficult because its value:
\begin{itemize}
    \item Is influenced by temperature, electric field, etc. which make it difficult to isolate the parameter;
    \item Varies depending on the defect type and interaction. For a point defect with several charge states, the cross-section can vary strongly depending on the strength of the Coulomb interaction;
    \item Is measured indirectly through fitting experimental data.
\end{itemize}
This non-exhaustive list of challenges has resulted in an immense amount of uncertainty with a diverse spectrum of reported values that vary approximately 7-9 orders of magnitude with the average cross-section being between $10^{-12}-10^{-18}$ cm$^2$.

In GaAs, radiation-induced defects (Fig. \ref{fig:fig2}) exist in positive, neutral, and negative charge states. To mimic the different Coulomb interactions between each charge state and carrier \cite{roosbroeck1950}, the cross-section is varied between the average experimental values of $10^{-12}-10^{-18}$~cm$^2$. We have systematically chosen $1.0\times10^{-12}$, $1.0\times10^{-15}$, and $1.0\times10^{-18}$ cm$^{2}$ for attractive (Reactions 1 and 4), neutral (capture by a neutral defect, e.g., Reactions 2 and 3), and repulsive (e.g., $a^{1+}+h^+\rightleftharpoons a^{2+}$) interactions, respectively. Chosen values are in range of experiments \cite{taghizadeh2018} and \cite{Benton_1982}, are of the correct order, and reflect the varying Coulomb interactions.

\subsection{Diffusion Migration Barrier}
The diffusion of defects in a bulk semiconductor is a challenge for both experimental and theoretical methods. Interstitials often possess complex diffusion pathways that can be highly temperature-dependent and result in chemical reactions with other defects. As a result, direct measurements are typically nonexistent, mandating greater reliance upon DFT calculations to obtain defect diffusion parameters. 

In GaAs, vacancies are immobile and interstitials are mobile. The diffusion coefficient that encompasses the migration barrier was chosen to be 0~cm/s$^2$ for immobile vacancies. The mobile $As_i$ and $Ga_i$ were calculated using DFT. Experiment has inferred~\cite{Bourg88} and DFT calculations have confirmed~\cite{PAS09} that As$_i^{1+}$ has a small thermal migration barrier and will also diffuse athermally via dthe Bourgoin-Corbett mechanism \cite{bourgoin-corbett}. However, the focus in this work is the $Ga_i$ which has a much higher barrier ($E_{m}^{DFT}\approx1.0$~eV~\cite{schultz2009}), is not athermal, and has also not (yet) been observed by experiment. The migration barrier is a key diffusion parameter and must be carefully considered. Decreasing the barrier by 0.1 eV can accelerate the rate of the defect-defect reaction by two orders of magnitude. This may not be an issue if the investigation is time-dependent. The migration barriers for all defect species are provided in Table~\ref{tab:defectpops}. 

\subsection{Effective Radius}
The effective radius R$_{eff}$ (Eqn. \ref{eqn:ddreact}) is a computational quantity used to describe the spatial range within which a defect will diffuse and react with another defect. It represents a simplified model---proxy---for physical and chemical interactions between defects, i.e., a smaller radius---shorter distance---indicates a faster reaction rate. The value of the effective radius is drawn from the Onsager radius which for semiconductors is typically on the order of the lattice constant (for GaAs, R$_{eff}\approx10^{-8}$~cm). However, the Onsager radius and, in general, collision theory treats species like point charges (hard spheres) and neglects long-range forces, quantum effects, and any molecular complexity, i.e., how the species must arrange themselves to interact. As a result, without an alternative, an effective radius of approximately the lattice constant is a viable quantity to describe the range of interaction.

The effective radius can also be used to mimic the Coulomb attraction between species and will certainly be larger (smaller) for more (less) attracted species. As a result of the limited information in the effective radius, we find that the reaction radius can vary by 1-2 orders of magnitude from the lattice constant, depending on the interaction between the two defects. Therefore, the R$_{eff}$ values for neutral and charged interactions ($A^0+B^{1+}$) should be similar to the material lattice parameter; for example, for GaAs, R$_{eff}^{GaAs}\approx5.0\times10^{-8}$~cm. For charged attractive interactions ($A^{1+}+B^{1-}$), R$_{eff}$ should be larger than the material lattice constant. Increasing (decreasing) the effective radius will increase (decrease) the rate of the reaction, the amount depends on the other parameters that are included in the simulation. 

\section{Results and Discussion}
\label{sec:results}
Electron irradiation of Si-doped GaAs generates displacement damage, resulting in the formation of interstitials ($Ga_i$ and $As_i$) and vacancies ($v_{As}$, $v_{Ga}$, $v_{Ga}v_{As}$). To showcase the capabilities of the AIDE method, we simulate the dynamical processes associated with these defects including Fermi level movement (Sec. \ref{subsec:fermi}), charge equilibration (Sec. \ref{subsec:ce}), and a defect-defect reaction (Sec.~\ref{subsec:diffusion}). The simulation space is composed of a simple one-dimensional slab of length $0.21\times10^{-4}$~cm with two contacts and boundary conditions that do not allow defects to diffuse through the contact barriers. The REOS band gap was populated with the DFT band gap (1.54 \cite{schultz2009}) and defect levels (Fig. \ref{fig:fig2}). Their populations are provided in Table \ref{tab:defectpops}. Other experimental parameter values used in this work are provided in Table \ref{tab:exp.params}. Note, Secs. \ref{subsec:fermi} and \ref{subsec:ce} focus primarily on electron-hole dynamics with ions fixed in the lattice. In Sec. \ref{subsec:diffusion}, ionic dynamics is explored.

\begin{table}[h!]
\renewcommand{\arraystretch}{1.3}
\caption{\label{tab:exp.params} Experimental parameters used in the REOS suite.}
\centering

\begin{tabularx}{\columnwidth}{lc}
\hline
\bfseries Parameter&
\bfseries Value\\ \hline \hline
Band gap ($E_g$) & 1.52 eV at 0K \cite{vurgaftman2001} \\
Dielectric Constant ($\varepsilon$) & 10.88 \cite{blakemore1982} \\
Electron Diffusion Coefficient $D_e$ & 220 \cite{neuberger} \\
Electron Effective Mass $m_e^*$ & 0.067 $m_e$\\
Hole Diffusion Coefficient $D_h$ & 10 \cite{neuberger} \\
Hole Effective Mass $m_h^*$ & 0.45 $m_e$ \\
Intrinsic carrier concentration ($n_i$) & $2.0\times10^6$ cm$^{-3}$ \\
Temperature (T) & 300 K \\
\hline
\end{tabularx}
\end{table}

\subsection{Fermi Level Position}
\label{subsec:fermi}
The Fermi level ($E_F$) location is a fundamental characteristic of any semiconductor material and device, providing insight into a material’s electronic properties including which defect charge states are most populated. An irradiation event disrupts the distribution of electrons and holes, causing the Fermi level to split into their respective quasi-Fermi levels. After the irradiation event, charge flows throughout the system until a new equilibrium is achieved---the quasi-Fermi levels converge into a single equilibrium Fermi level. To demonstrate this capability, $e^-$ and $h^+$ capture and emission reactions are performed in REOS on the irradiated Si-doped GaAs sample. It is not feasible to know the exact population of each defect charge state throughout the irradiation process. As a result, an initial population (Table~\ref{tab:defectpops}) is given to each neutral charge state and is allowed to distribute itself into the system---charge equilibration---until equilibrium is achieved. Charge carrier reactions drive the charge flow in the material and dictate where the Fermi level will stabilize.  

\begin{figure}[t]
    \centering
    \includegraphics[width=\columnwidth]{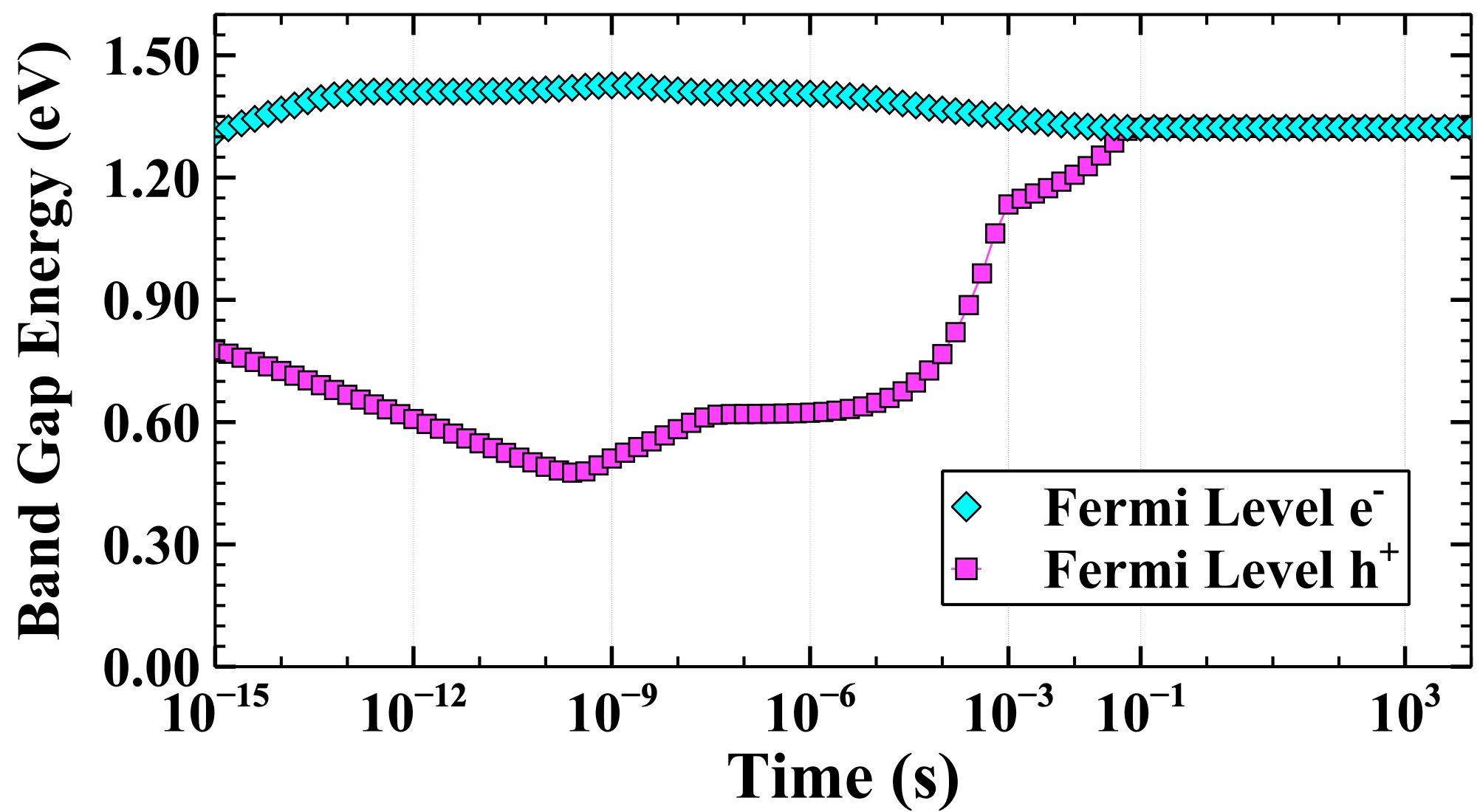}
    \caption{REOS computed Fermi level for the $e^-$ and $h^+$ in GaAs. Equilibrium is achieved in roughly 1 s where the Fermi level for the charge carriers reconcile at $E_F\approx1.32$ eV.}
    \label{fig:fig4}
\end{figure}

In Fig. \ref{fig:fig4}, the chemical potentials---quasi-Fermi levels (Eqns. \ref{eqn:fermie} and \ref{eqn:fermih})---for the $e^-$ and $h^+$ are shown with equilibrium occurring in about $10^{-1}$ s. Initially, the $e^-$ Fermi level increases towards the doping level ($\approx$1.52 eV above the valence band edge) as $e^-$ ($h^+$) are released (captured)--Si-dopants release electrons--but sinks into the band gap as electrons are captured by deeper defects. The Fermi levels of the charge carriers are initially dissimilar and remain in quasi-equilibrium until charge equilibration is complete, resulting in stabilization, and convergence at $E_F\approx1.32$ eV. Small fluctuations in the Fermi levels are due to the changing defect population sizes as electrons are being captured. This process can be laborious because to reach an $E_F\approx1.32$ eV, charge has to fill several states for each defect. If a lower doping concentration were used, the Fermi level would be lower, there would be less states to fill, and the quasi-Fermi level convergence process would occur faster. This capability can be highly beneficial to recent works which have utilized the concept of quasi-Fermi levels for many systems including semiconducting devices and alloys \cite{loper2011,jain2022,Reddy2017,Riley2021}, photovoltaics \cite{Lombez2014} and solar cells \cite{phuong2021,warby2023}, and two-dimensional field-effect transistors \cite{yan2022}.

\subsection{Charge Equilibration}
\label{subsec:ce}
\begin{figure}[t!]
    \centering
    \includegraphics[width=\columnwidth]{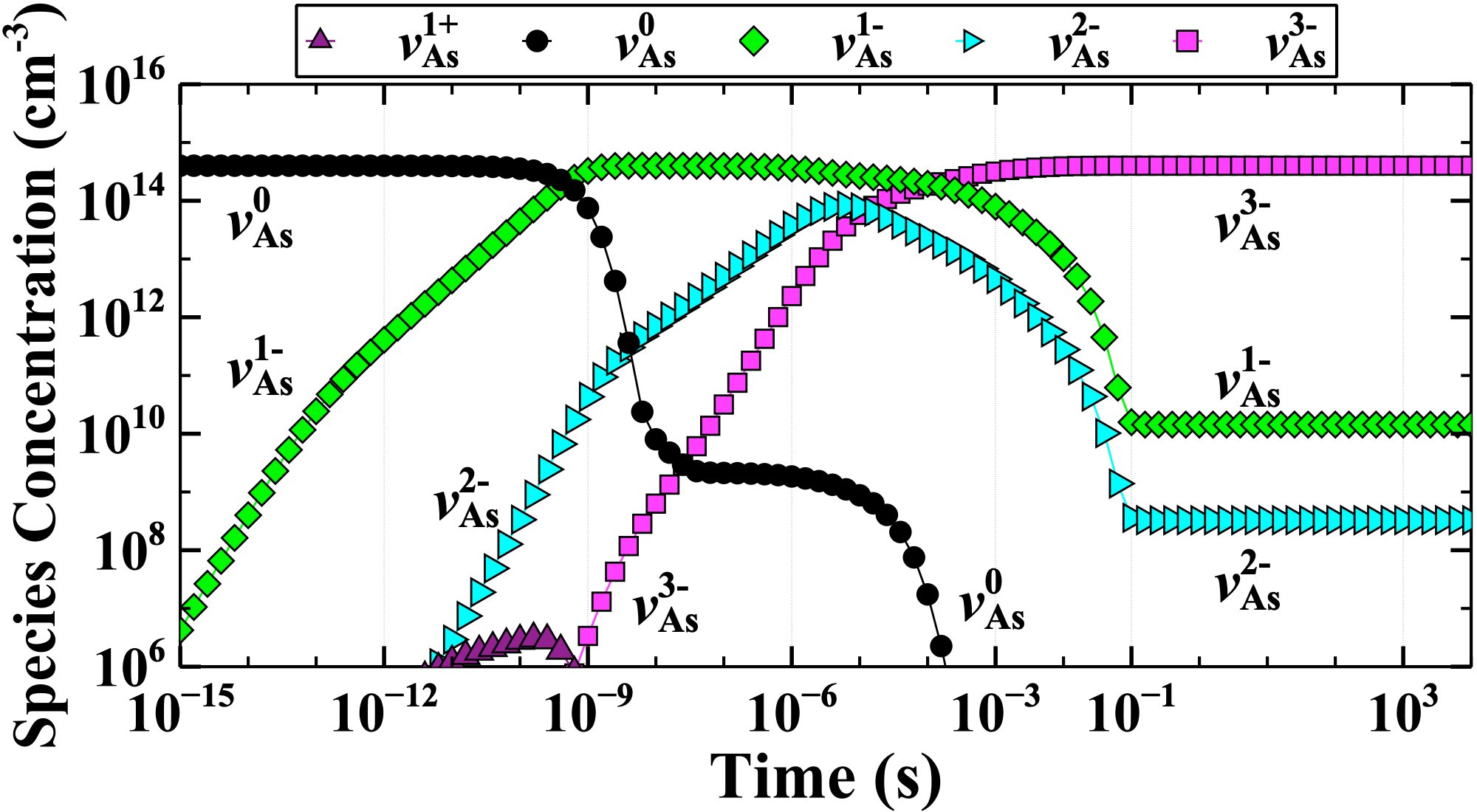}
    \caption{Calculated defect species density of $v_{As}$ during charge equilibration. All charge states were included but $v_{As}^{2+}$ and $v_{As}^{3+}$ are well below $10^6$ cm$^{-3}$ and are not shown in the figure. The most populated charge state is $v_{As}^{3-}$ at slightly less than $1.0\times10^{15}$ cm$^{-3}$.}
    \label{fig:fig5}
\end{figure}

To verify the Fermi level shift, species populations were investigated for each defect. Although all defects were included in the simulation, we focus our discussion on $v_{As}$. In Fig.~\ref{fig:fig5}, the species density of the arsenic vacancy in all its charge states is shown. In Fig.~\ref{fig:fig5}, the evolution of the As vacancy charge state populations is shown. The evolution seems complicated. However, when compared with the defect levels in Fig.~\ref{fig:fig2} it becomes apparent that charge states are being filled, from the valence band towards the conduction band, by dopant-electrons until stabilization is achieved and the highest populated charge state emerges---the $v_{As}^{3-}$ is the most populated. Per our model, charge equilibration begins before $10^{-15}$~s with all defects, except the neutral $v_{As}$ having densities well below experimental probing limits. As the charge distributes between the charge states $v_{As}^{3-}$ emerges as the dominant species settling with a population of $\approx1.0\times10^{14}$~cm$^{-3}$ after $\approx10^{-3}$~s. This $-3$ charge state verifies the Fermi level position because, as shown in Fig.~\ref{fig:fig2}, at $E_F\approx1.32$~eV the $v_{As}^{3-}$ is the "surviving" species for the As vacancy, i.e., a vast majority of the As vacancy exists in the $v_{As}^{3-}$. Notably, the model correctly identified the defect species with the largest population sizes for all defect types, including defects that exhibit negative U behavior (i.e. $v_{As}$).

In addition to correctly identifying $v_{As}^{3-}$ as the dominant $v_{As}$ charge state, the approach also predicts the charge states for the other defect species, including $As_{i}^{1-}$, $Ga_i^{1+}$, $v_{Ga}^{3-}$, and $v_{Ga}v_{As}^{2-}$. This consistency between defect levels and the Fermi level serves as a metric for the accuracy of the model and AIDE's potential to perform virtual experiments to access the accuracy of (validate) defect levels.

Charge equilibration is critical to dynamically modeling defects and significantly impacts the interactions between defects. The AIDE approach allows one to predict the possible defect chemistry at different Fermi levels, doping concentrations, and damage levels.

\subsection{Diffusion-Driven Defect Reactions}
\label{subsec:diffusion}
Though DFT can accurately describe defect chemistry in a static environment, it is limited in simulating any dynamical behavior. Therefore, another goal of this defect-informed multiscale approach is to use parameters from the atomistic realm and available experimental measurements to simulate defect-defect chemistry on a continuum scale. This entails using approximated data and filling gaps with activation energies and diffusion migration energies from DFT. Limiting this study to diffusion-driven reactions, the difficult to determine diffusion prefactor and migration barrier are the parameters driving the reaction (Eqn. \ref{eqn:diffusion}). To study this capability, we showcase the following reaction:

\begin{reactions*}
    \hspace{.4cm}&\llap{Reaction 6} & \emph{v}_{As}^{3-}  + $Ga$_{i}^{1+} & <=>  $Ga$_{As}^{2-}
\end{reactions*}
where the $Ga_i$ can jump into an $v_{As}$ creating a gallium antisite (gallium atom occupying an open arsenic site) in a $-2$ charge state. As a prerequisite to our model, the $Ga_{As}^{2-}$ was shown to be a stable defect charge state by DFT calculations \cite{schultz2009} and is shown in Fig.~\ref{fig:fig2}. This reaction was chosen for two reasons: firstly, the $Ga_i$ is thermally mobile and can find the immobile $v_{As}$ (Table~\ref{tab:defectpops}); secondly, the two reacting defects are oppositely charged and experience a Coulomb attraction. To account for this extended range of Coulomb attraction, the effective radius was set to R$_{eff}=1.0\times10^{-6}$ cm---larger than the lattice parameter R$_{eff}\approx5.0\times10^{-8}$ cm. 

\begin{figure}[t!]
    \centering
    \includegraphics[width=\columnwidth]{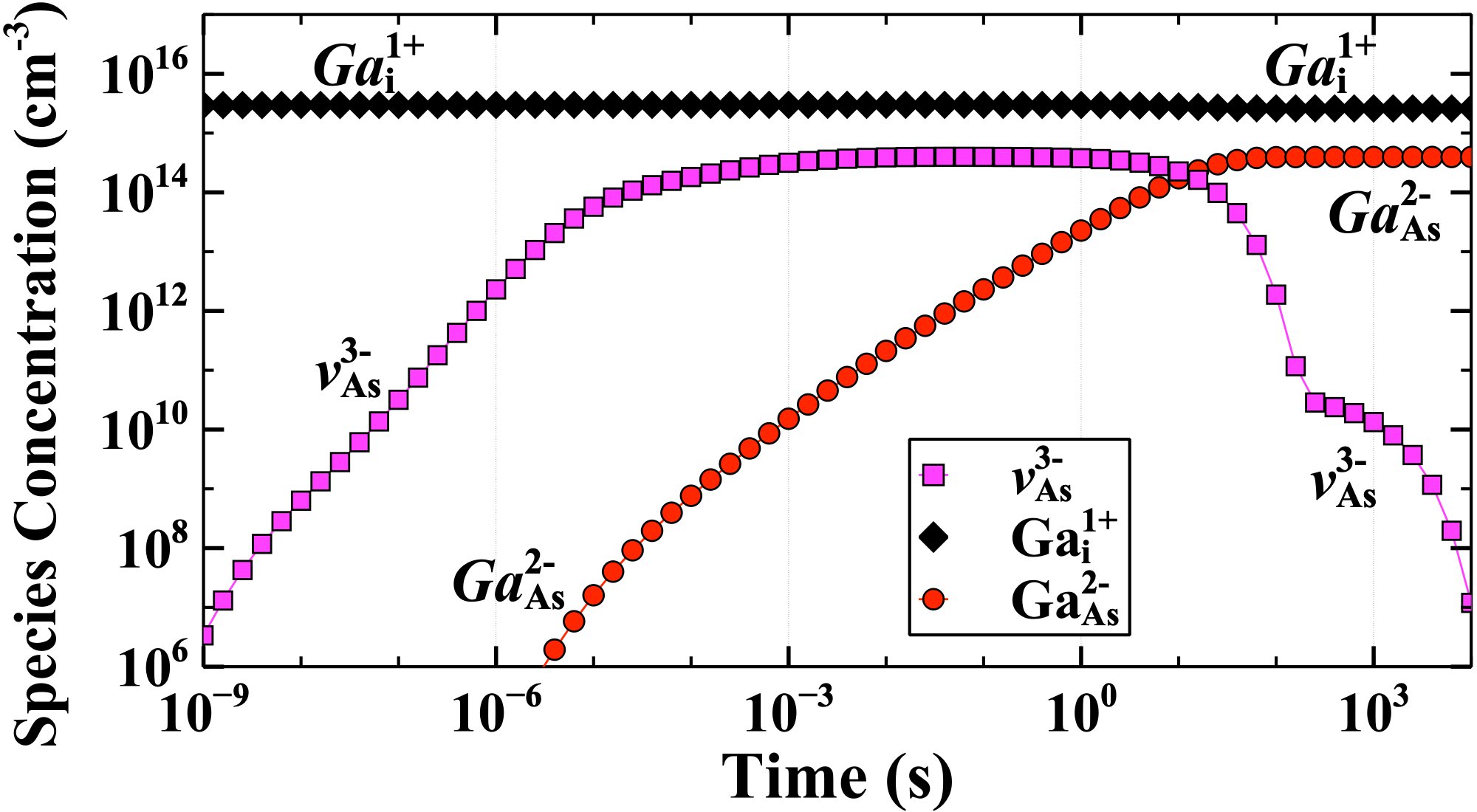}
    \caption{Diffusion-limited defect-defect reaction between $v_{As}^{1-}$ and $Ga_i^{1+}$. The formation of $Ga_{As}^{2-}$ grows from 0 to roughly $10^{14}$ cm$^{-3}$ over an extremely long period of time.}
    \label{fig:fig6}
\end{figure}

In Fig.~\ref{fig:fig6}, the reaction between the two defects and their resulting product are shown (Eqns. \ref{eqn:diffusion}, \ref{eqn:ddrate}, and \ref{eqn:ddreact}). During the initial stages of the reaction, the $v_{As}^{3-}$ population is small ($<10^6$~cm$^{-3}$) causing the reaction product ($Ga_{As}^{2-}$) to grow at a slow rate. This small $v_{As}^{3-}$ population is due to charge equilibration, i.e., dopant electrons filling charge states below it in the band gap. Note, this process occurs also for the $Ga_i$ but occurs much faster and is completed before $10^{-9}$~s. After approximately $10^{-4}$~s, the $v_{As}^{3-}$ reaches its peak and the $Ga_{As}^{2-}$ starts growing with a constant slope until reaching its maximum concentration. The growing $Ga_{As}^{2-}$ population peaks after $10^1$~s where its concentration remains fairly constant along with the population of $Ga_i^{1+}$. Since the product formation is limited by the reactant with a smaller population size, the $Ga_{As}^{2-}$ peaks when the $v_{As}^{3-}$ population decays to less than $10^6$~cm$^{-3}$. The decay occurs because the larger $Ga_i^{1+}$ population largely consumes the available $v_{As}^{3-}$. 


The $Ga_{As}^{2-}$ achieving its maximum population in $10^1$~s is likely a nonphysical length of time for such a simple reaction. The cause of this lengthy reaction can be explained by the diffusion of the Ga interstitial. The $Ga_i^{1+}$ has a thermal migration barrier of $\approx1.0$~eV~\cite{schultz2009}, and is a slow diffuser relative to the fast athermal As interstitial whose barrier has been experimentally reported (and theoretically validated) as $E_m^{As^{1+}}=0.5$~eV~\cite{bourgoin1988,schultz2009}. Unfortunately, no experimental observation of the $Ga_i$ or its diffusion properties, including its migration barrier, have been reported. We must resort to the DFT-computed values. Despite this limitation, we have demonstrated a dynamical chemical reaction between two Coulomb-attracted defects, and any additional experimental measurements will lead to a more physically accurate result. 

Notably, several other defect-defect reactions are expected to occur in GaAs. These reactions can also be implemented into the REOS simulation suite, which is beyond the scope of this work and will be presented in follow-on work. 

\section{Conclusion}
\label{sec:conclusion}
Using the REOS suite, we have demonstrated a defect-aware model for dynamically simulating the defect chemistry in irradiated Si-doped GaAs. The AIDE method successfully predicts the Fermi level behavior through electron/hole reactions. The quasi-Fermi levels varied considerably, but reconciled when equilibrium was achieved. The Fermi level position was verified by monitoring the most populated charge state for the As vacancy, i.e., the $v_{As}^{3-}$ was found to be the dominant species. Additionally, a defect-defect reaction successfully results in the formation of $Ga_{As}^{2-}$. The details of this dynamical process illustrated the rich insight gained by studying the reaction process with the AIDE method.  

The AIDE method is a generalizable approach to studying defect chemistry in semiconducting materials. Extending the AIDE approach to other semiconducting systems depends on the availability of sufficiently accurate defect levels, diffusion migration barriers, etc. from DFT and/or experimental measurements. For experimentally and theoretically challenging materials like InGaAs, virtual experiments can be performed in REOS to understand, quantify, and bound relevant and difficult-to-determine physical quantities (e.g., capture cross-section, migration barriers, etc.). However, prior to being applied to InGaAs, this atoms-to-devices approach will be used to examine remaining open questions in GaAs such as characterizing the remaining unidentified DLTS peaks and, perhaps more interesting, the cause of the Fermi level pinning~\cite{Walukiewicz1988}. 

To validate this approach, further experimental studies are needed. One experiment, in particular, would provide insight into the amount or level of displacement damage with respect to doping concentration. Despite several DLTS studies, direct discussion of the level of damage is minimal and incomplete, adding uncertainty to theoretical studies. Another experiment could provide information on the diffusion properties of the Ga interstitial. Previous DFT calculations suggest that the Ga interstitial has a large migration barrier of $\approx1.0$~eV~\cite{schultz2009} but with no direct experimental observation of an interstitial in GaAs, or (nearly) any other III-V material, the reliability of such computations will always be in question. Perhaps an experiment could probe the unexplained 235~K annealing stage~\cite{thommen1970} that might be the Ga interstitial. These types of experimental studies could provide justification for our approximations and enable us to improve our model. 

The AIDE method is shown to be a powerful tool for simulating defect chemistry and providing crucial insight into defect annealing behavior both at experimentally accessible and inaccessible regimes. With the AIDE method, the REOS suite will serve as an arena to perform virtual experiments on relevant quantities in defect/device modeling. 

\section*{Acknowledgment}
We thank G. Vizkelethy, J. M. Cain, B. A. Aguirre, E. S. Bielejec, and P. J. Griffin for useful discussion. We are indebted to D. Black and J. Black for comments on the manuscript. Sandia National Laboratories is a multimission laboratory managed and operated by National Technology \& Engineering Solutions of Sandia, LLC, a wholly owned subsidiary of Honeywell International Inc., for the U.S. Department of Energy’s National Nuclear Security Administration under contract No. DE-NA0003525. This work was supported by a Laboratory Directed Research and Development (LDRD) project (No. 229430, SAND2025-08646PE).

\bibliography{aps}

\providecommand{\noopsort}[1]{}\providecommand{\singleletter}[1]{#1}%
\begin{thebibliography}{57}%
\makeatletter
\providecommand \@ifxundefined [1]{%
 \@ifx{#1\undefined}
}%
\providecommand \@ifnum [1]{%
 \ifnum #1\expandafter \@firstoftwo
 \else \expandafter \@secondoftwo
 \fi
}%
\providecommand \@ifx [1]{%
 \ifx #1\expandafter \@firstoftwo
 \else \expandafter \@secondoftwo
 \fi
}%
\providecommand \natexlab [1]{#1}%
\providecommand \enquote  [1]{``#1''}%
\providecommand \bibnamefont  [1]{#1}%
\providecommand \bibfnamefont [1]{#1}%
\providecommand \citenamefont [1]{#1}%
\providecommand \href@noop [0]{\@secondoftwo}%
\providecommand \href [0]{\begingroup \@sanitize@url \@href}%
\providecommand \@href[1]{\@@startlink{#1}\@@href}%
\providecommand \@@href[1]{\endgroup#1\@@endlink}%
\providecommand \@sanitize@url [0]{\catcode `\\12\catcode `\$12\catcode
  `\&12\catcode `\#12\catcode `\^12\catcode `\_12\catcode `\%12\relax}%
\providecommand \@@startlink[1]{}%
\providecommand \@@endlink[0]{}%
\providecommand \url  [0]{\begingroup\@sanitize@url \@url }%
\providecommand \@url [1]{\endgroup\@href {#1}{\urlprefix }}%
\providecommand \urlprefix  [0]{URL }%
\providecommand \Eprint [0]{\href }%
\providecommand \doibase [0]{https://doi.org/}%
\providecommand \selectlanguage [0]{\@gobble}%
\providecommand \bibinfo  [0]{\@secondoftwo}%
\providecommand \bibfield  [0]{\@secondoftwo}%
\providecommand \translation [1]{[#1]}%
\providecommand \BibitemOpen [0]{}%
\providecommand \bibitemStop [0]{}%
\providecommand \bibitemNoStop [0]{.\EOS\space}%
\providecommand \EOS [0]{\spacefactor3000\relax}%
\providecommand \BibitemShut  [1]{\csname bibitem#1\endcsname}%
\let\auto@bib@innerbib\@empty
\bibitem [{\citenamefont {{R. J. Walters, G. J. Shaw, G. P. Summers, E. A.
  Burke, and S. R. Messenger}}(1992)}]{walters1992}%
  \BibitemOpen
  \bibfield  {author} {\bibinfo {author} {\bibnamefont {{R. J. Walters, G. J.
  Shaw, G. P. Summers, E. A. Burke, and S. R. Messenger}}},\ }\bibfield
  {title} {\bibinfo {title} {{Radiation effects in Ga$_{0.47}$In$_{0.53}$As
  devices}},\ }\href {https://doi.org/10.1109/23.211429} {\bibfield  {journal}
  {\bibinfo  {journal} {IEEE Trans. Nuc. Sci.}\ }\textbf {\bibinfo {volume}
  {39}},\ \bibinfo {pages} {2257} (\bibinfo {year} {1992})}\BibitemShut
  {NoStop}%
\bibitem [{\citenamefont {{G. J. Shaw, S. R. Messenger, R. J. Walters, and G.
  P. Summers}}(1993{\natexlab{a}})}]{shaw1993}%
  \BibitemOpen
  \bibfield  {author} {\bibinfo {author} {\bibnamefont {{G. J. Shaw, S. R.
  Messenger, R. J. Walters, and G. P. Summers}}},\ }\bibfield  {title}
  {\bibinfo {title} {{Radiation‐induced reverse dark currents in
  In$_{0.53}$Ga$_{0.47}$As photodiodes}},\ }\href
  {https://doi.org/https://doi.org/10.1063/1.354012} {\bibfield  {journal}
  {\bibinfo  {journal} {J. Appl. Phys.}\ }\textbf {\bibinfo {volume} {73}},\
  \bibinfo {pages} {7244} (\bibinfo {year} {1993}{\natexlab{a}})}\BibitemShut
  {NoStop}%
\bibitem [{\citenamefont {{G. J. Shaw, S. R. Messenger, R. J. Walters, and G.
  P. Summers}}(1993{\natexlab{b}})}]{shaw1993_2}%
  \BibitemOpen
  \bibfield  {author} {\bibinfo {author} {\bibnamefont {{G. J. Shaw, S. R.
  Messenger, R. J. Walters, and G. P. Summers}}},\ }\bibfield  {title}
  {\bibinfo {title} {{Time dependence of radiation‐induced generation
  currents in irradiated InGaAs photodiodes}},\ }\href
  {https://doi.org/https://doi.org/10.1063/1.354812} {\bibfield  {journal}
  {\bibinfo  {journal} {J. Appl. Phys.}\ }\textbf {\bibinfo {volume} {74}},\
  \bibinfo {pages} {1629} (\bibinfo {year} {1993}{\natexlab{b}})}\BibitemShut
  {NoStop}%
\bibitem [{\citenamefont {{X. Hugon, O. Amore, S. Cortial, C. Lenoble, and M.
  Villard}}(1995)}]{hugon1995}%
  \BibitemOpen
  \bibfield  {author} {\bibinfo {author} {\bibnamefont {{X. Hugon, O. Amore, S.
  Cortial, C. Lenoble, and M. Villard}}},\ }\bibfield  {title} {\bibinfo
  {title} {{Near-room operating temperature SWIR InGaAs detectors in
  progress}},\ }\href {https://doi.org/https://doi.org/10.1117/12.218273}
  {\bibfield  {journal} {\bibinfo  {journal} {Proc. SPIE, B. F. Andresen and M.
  Strojnik}\ }\textbf {\bibinfo {volume} {2552}},\ \bibinfo {pages} {738}
  (\bibinfo {year} {1995})}\BibitemShut {NoStop}%
\bibitem [{\citenamefont {{S. Barde, R. Ecoffet, J. Costeraste, A. Meygret, and
  X. Hugon}}(2000)}]{barde2000}%
  \BibitemOpen
  \bibfield  {author} {\bibinfo {author} {\bibnamefont {{S. Barde, R. Ecoffet,
  J. Costeraste, A. Meygret, and X. Hugon}}},\ }\bibfield  {title} {\bibinfo
  {title} {{Displacement damage effects in InGaAs detectors: experimental
  results and semi-empirical model prediction}},\ }\href
  {https://doi.org/10.1109/23.903794} {\bibfield  {journal} {\bibinfo
  {journal} {IEEE Trans. Nuc. Sci.}\ }\textbf {\bibinfo {volume} {47}},\
  \bibinfo {pages} {2466} (\bibinfo {year} {2000})}\BibitemShut {NoStop}%
\bibitem [{\citenamefont {{O. Gilard, L. S. How, A. Delbergue, C. Inguimbert,
  T. Nuns, J. Barbero, J. Moreno, L. Bouet, S. Mariojouls, and M.
  Boutillier}}(2018)}]{gilard2018}%
  \BibitemOpen
  \bibfield  {author} {\bibinfo {author} {\bibnamefont {{O. Gilard, L. S. How,
  A. Delbergue, C. Inguimbert, T. Nuns, J. Barbero, J. Moreno, L. Bouet, S.
  Mariojouls, and M. Boutillier}}},\ }\bibfield  {title} {\bibinfo {title}
  {{Damage Factor for Radiation-Induced Dark Current in InGaAs Photodiodes}},\
  }\href {https://doi.org/10.1109/TNS.2018.2799742} {\bibfield  {journal}
  {\bibinfo  {journal} {IEEE Trans. Nuc. Sci.}\ }\textbf {\bibinfo {volume}
  {65}},\ \bibinfo {pages} {884} (\bibinfo {year} {2018})}\BibitemShut
  {NoStop}%
\bibitem [{\citenamefont {{T. Nuns, C. Inguimbert, J. Barbero, J. Moreno, S.
  Ducret, A. Nedelcu, B. Galnander, and E. Passoth}}(2020)}]{nuns2020}%
  \BibitemOpen
  \bibfield  {author} {\bibinfo {author} {\bibnamefont {{T. Nuns, C.
  Inguimbert, J. Barbero, J. Moreno, S. Ducret, A. Nedelcu, B. Galnander, and
  E. Passoth}}},\ }\bibfield  {title} {\bibinfo {title} {{Displacement Damage
  Effects in InGaAs Photodiodes due to Electron, Proton, and Neutron
  Irradiation}},\ }\href {https://doi.org/10.1109/TNS.2020.2984133} {\bibfield
  {journal} {\bibinfo  {journal} {IEEE Trans. Nuc. Sci.}\ }\textbf {\bibinfo
  {volume} {67}},\ \bibinfo {pages} {1263} (\bibinfo {year}
  {2020})}\BibitemShut {NoStop}%
\bibitem [{\citenamefont {{G. T. Nelson, G. Ouin, S. J. Polly, K. B. Wynne, A.
  W. Haberl, W. A. Lanford, R. A. Lowell}}(2020)}]{nelson2020}%
  \BibitemOpen
  \bibfield  {author} {\bibinfo {author} {\bibnamefont {{G. T. Nelson, G. Ouin,
  S. J. Polly, K. B. Wynne, A. W. Haberl, W. A. Lanford, R. A. Lowell}}},\
  }\bibfield  {title} {\bibinfo {title} {{In Situ Deep-level Transient
  Spectroscopy and Dark Current Measurements of Proton-Irradiated InGaAs}},\
  }\href {https://doi.org/10.1109/TNS.2020.3011729} {\bibfield  {journal}
  {\bibinfo  {journal} {IEEE Trans. Nuc. Sci.}\ }\textbf {\bibinfo {volume}
  {67}},\ \bibinfo {pages} {2051} (\bibinfo {year} {2020})}\BibitemShut
  {NoStop}%
\bibitem [{\citenamefont {{J. C. Bourgoin, H. J. von Bardeleben, and D.
  Stiévenard}}(1988)}]{bourgoin1988}%
  \BibitemOpen
  \bibfield  {author} {\bibinfo {author} {\bibnamefont {{J. C. Bourgoin, H. J.
  von Bardeleben, and D. Stiévenard}}},\ }\bibfield  {title} {\bibinfo {title}
  {{Native defects in gallium arsenide}},\ }\href
  {https://doi.org/10.1063/1.341206} {\bibfield  {journal} {\bibinfo  {journal}
  {J. Appl. Phys.}\ }\textbf {\bibinfo {volume} {64}},\ \bibinfo {pages} {R65}
  (\bibinfo {year} {1988})}\BibitemShut {NoStop}%
\bibitem [{\citenamefont {{D. Pons and J. C.
  Bourgoin}}(1985)}]{pons_bourgoin1985}%
  \BibitemOpen
  \bibfield  {author} {\bibinfo {author} {\bibnamefont {{D. Pons and J. C.
  Bourgoin}}},\ }\bibfield  {title} {\bibinfo {title} {{Irradiation-induced
  defects in GaAs}},\ }\href {https://doi.org/10.1088/0022-3719/18/20/012}
  {\bibfield  {journal} {\bibinfo  {journal} {J. Phys. C: Sol. State Phys.}\
  }\textbf {\bibinfo {volume} {18}},\ \bibinfo {pages} {3839} (\bibinfo {year}
  {1985})}\BibitemShut {NoStop}%
\bibitem [{\citenamefont {{D. Stievenard, X. Boddaert, J. C. Bourgoin, and H.
  J. von Bardeleben}}(1990)}]{Stievenard1990}%
  \BibitemOpen
  \bibfield  {author} {\bibinfo {author} {\bibnamefont {{D. Stievenard, X.
  Boddaert, J. C. Bourgoin, and H. J. von Bardeleben}}},\ }\bibfield  {title}
  {\bibinfo {title} {{Behavior of electron-irradiation-induced defects in
  GaAs}},\ }\href {https://doi.org/10.1103/PhysRevB.41.5271} {\bibfield
  {journal} {\bibinfo  {journal} {Phys. Rev. B}\ }\textbf {\bibinfo {volume}
  {41}},\ \bibinfo {pages} {5271} (\bibinfo {year} {1990})}\BibitemShut
  {NoStop}%
\bibitem [{\citenamefont {{G. A. Baraff and M.
  Schl\"uter}}(1985)}]{baraff1985}%
  \BibitemOpen
  \bibfield  {author} {\bibinfo {author} {\bibnamefont {{G. A. Baraff and M.
  Schl\"uter}}},\ }\bibfield  {title} {\bibinfo {title} {{Electronic Structure,
  Total Energies, and Abundances of the Elementary Point Defects in GaAs}},\
  }\href {https://doi.org/10.1103/PhysRevLett.55.1327} {\bibfield  {journal}
  {\bibinfo  {journal} {Phys. Rev. Lett.}\ }\textbf {\bibinfo {volume} {55}},\
  \bibinfo {pages} {1327} (\bibinfo {year} {1985})}\BibitemShut {NoStop}%
\bibitem [{\citenamefont {{S. B. Zhang and J. E. Northrup}}(1991)}]{zhang1991}%
  \BibitemOpen
  \bibfield  {author} {\bibinfo {author} {\bibnamefont {{S. B. Zhang and J. E.
  Northrup}}},\ }\bibfield  {title} {\bibinfo {title} {{Chemical potential
  dependence of defect formation energies in GaAs: application to Ga
  self-diffusion}},\ }\href {https://doi.org/10.1103/PhysRevLett.67.2339}
  {\bibfield  {journal} {\bibinfo  {journal} {Phys. Rev. Lett.}\ }\textbf
  {\bibinfo {volume} {67}},\ \bibinfo {pages} {2339} (\bibinfo {year}
  {1991})}\BibitemShut {NoStop}%
\bibitem [{\citenamefont {{P. A. Schultz and O. Anatole von
  Lilienfeld}}(2009)}]{schultz2009}%
  \BibitemOpen
  \bibfield  {author} {\bibinfo {author} {\bibnamefont {{P. A. Schultz and O.
  Anatole von Lilienfeld}}},\ }\bibfield  {title} {\bibinfo {title} {{Simple
  intrinsic defects in gallium arsenide}},\ }\href
  {https://doi.org/10.1088/0965-0393/17/8/084007} {\bibfield  {journal}
  {\bibinfo  {journal} {Modelling Simul. Mater. Sci. Eng.}\ }\textbf {\bibinfo
  {volume} {17}},\ \bibinfo {pages} {084007} (\bibinfo {year}
  {2009})}\BibitemShut {NoStop}%
\bibitem [{\citenamefont {Schultz}(2015)}]{schultz2015}%
  \BibitemOpen
  \bibfield  {author} {\bibinfo {author} {\bibfnamefont {P.~A.}\ \bibnamefont
  {Schultz}},\ }\bibfield  {title} {\bibinfo {title} {{The E1–E2 center in
  gallium arsenide is the divacancy}},\ }\href
  {https://doi.org/10.1088/0953-8984/27/7/075801} {\bibfield  {journal}
  {\bibinfo  {journal} {{J. Phys.: Cond. Mat.}}\ }\textbf {\bibinfo {volume}
  {27}},\ \bibinfo {pages} {075801} (\bibinfo {year} {2015})}\BibitemShut
  {NoStop}%
\bibitem [{\citenamefont {{P. A. Schultz and H. P.
  Hjalmarson}}(2022)}]{schultz2022}%
  \BibitemOpen
  \bibfield  {author} {\bibinfo {author} {\bibnamefont {{P. A. Schultz and H.
  P. Hjalmarson}}},\ }\bibfield  {title} {\bibinfo {title} {{Theory of the
  metastable injection-bleached $E3c$ center in GaAs}},\ }\href
  {https://doi.org/10.1103/PhysRevB.105.224111} {\bibfield  {journal} {\bibinfo
   {journal} {Phys. Rev. B}\ }\textbf {\bibinfo {volume} {105}},\ \bibinfo
  {pages} {224111} (\bibinfo {year} {2022})}\BibitemShut {NoStop}%
\bibitem [{\citenamefont {{F. Taghizadeh, F. Ostvar, F. D. Auret, and W. E.
  Meyer}}(2018)}]{taghizadeh2018}%
  \BibitemOpen
  \bibfield  {author} {\bibinfo {author} {\bibnamefont {{F. Taghizadeh, F.
  Ostvar, F. D. Auret, and W. E. Meyer}}},\ }\bibfield  {title} {\bibinfo
  {title} {{Laplace DLTS study of the fine structure and metastability of the
  radiation-induced E3 defect level in GaAs}},\ }\href
  {https://doi.org/10.1088/1361-6641/aae9a8} {\bibfield  {journal} {\bibinfo
  {journal} {Semicond. Sci. Technol.}\ }\textbf {\bibinfo {volume} {33}},\
  \bibinfo {pages} {125011} (\bibinfo {year} {2018})}\BibitemShut {NoStop}%
\bibitem [{\citenamefont {{P. A. Schultz}}(2016)}]{schultz2016}%
  \BibitemOpen
  \bibfield  {author} {\bibinfo {author} {\bibnamefont {{P. A. Schultz}}},\
  }\bibfield  {title} {\bibinfo {title} {{Discriminating a deep gallium
  antisite defect from shallow acceptors in GaAs using supercell
  calculations}},\ }\href {https://doi.org/10.1103/PhysRevB.93.125201}
  {\bibfield  {journal} {\bibinfo  {journal} {Phys. Rev. B}\ }\textbf {\bibinfo
  {volume} {93}},\ \bibinfo {pages} {125201} (\bibinfo {year}
  {2016})}\BibitemShut {NoStop}%
\bibitem [{\citenamefont {{F. El-Mellouhi and N.
  Mousseau}}(2006)}]{El-Mellouhi2006}%
  \BibitemOpen
  \bibfield  {author} {\bibinfo {author} {\bibnamefont {{F. El-Mellouhi and N.
  Mousseau}}},\ }\bibfield  {title} {\bibinfo {title} {{Charge-dependent
  migration pathways for the Ga vacancy in $\mathrm{GaAs}$}},\ }\href
  {https://doi.org/10.1103/PhysRevB.74.205207} {\bibfield  {journal} {\bibinfo
  {journal} {Phys. Rev. B}\ }\textbf {\bibinfo {volume} {74}},\ \bibinfo
  {pages} {205207} (\bibinfo {year} {2006})}\BibitemShut {NoStop}%
\bibitem [{\citenamefont {{W. Walukiewicz}}(1988)}]{Walukiewicz1988}%
  \BibitemOpen
  \bibfield  {author} {\bibinfo {author} {\bibnamefont {{W. Walukiewicz}}},\
  }\bibfield  {title} {\bibinfo {title} {{Mechanism of Fermi-level
  stabilization in semiconductors}},\ }\href
  {https://doi.org/10.1103/PhysRevB.37.4760} {\bibfield  {journal} {\bibinfo
  {journal} {Phys. Rev. B}\ }\textbf {\bibinfo {volume} {37}},\ \bibinfo
  {pages} {4760} (\bibinfo {year} {1988})}\BibitemShut {NoStop}%
\bibitem [{\citenamefont {{D. Colleoni and A.
  Pasquarello}}(2013)}]{colleoni2013}%
  \BibitemOpen
  \bibfield  {author} {\bibinfo {author} {\bibnamefont {{D. Colleoni and A.
  Pasquarello}}},\ }\bibfield  {title} {\bibinfo {title} {{Assignment of
  Fermi-level pinning and optical transitions to the (As$_{Ga}$)$_2$-O$_{As}$
  center in oxygen-doped GaAs}},\ }\href {https://doi.org/10.1063/1.4824309}
  {\bibfield  {journal} {\bibinfo  {journal} {{Appl. Phys. Lett.}}\ }\textbf
  {\bibinfo {volume} {103}},\ \bibinfo {pages} {142108} (\bibinfo {year}
  {2013})}\BibitemShut {NoStop}%
\bibitem [{\citenamefont {{D. Colleoni and A.
  Pasquarello}}(2016)}]{colleoni2016}%
  \BibitemOpen
  \bibfield  {author} {\bibinfo {author} {\bibnamefont {{D. Colleoni and A.
  Pasquarello}}},\ }\bibfield  {title} {\bibinfo {title} {{Oxygen defects in
  GaAs: A hybrid functional study}},\ }\href
  {https://doi.org/10.1103/PhysRevB.93.125208} {\bibfield  {journal} {\bibinfo
  {journal} {{Phys. Rev. B}}\ }\textbf {\bibinfo {volume} {93}},\ \bibinfo
  {pages} {125208} (\bibinfo {year} {2016})}\BibitemShut {NoStop}%
\bibitem [{\citenamefont {{X. Luo, J. Montes, S. D. Koukourinkova, B. L.
  Vaandrager, E. S. Bielejec, G. Vizkelethy, R. D. Schrimpf, D. M. Fleetwood,
  and E. X. Zhang}}(2024)}]{luo2024}%
  \BibitemOpen
  \bibfield  {author} {\bibinfo {author} {\bibnamefont {{X. Luo, J. Montes, S.
  D. Koukourinkova, B. L. Vaandrager, E. S. Bielejec, G. Vizkelethy, R. D.
  Schrimpf, D. M. Fleetwood, and E. X. Zhang}}},\ }\bibfield  {title} {\bibinfo
  {title} {{Low-frequency noise and deep level transient spectroscopy in n-p-n
  Si bipolar junction transistors irradiated with Si ions}},\ }\href
  {https://doi.org/10.1109/TNS.2023.3346834} {\bibfield  {journal} {\bibinfo
  {journal} {IEEE Trans. Nuc. Sci.}\ }\textbf {\bibinfo {volume} {71}},\
  \bibinfo {pages} {591} (\bibinfo {year} {2024})}\BibitemShut {NoStop}%
\bibitem [{\citenamefont {{D. V. Lang}}(1974)}]{lang1974}%
  \BibitemOpen
  \bibfield  {author} {\bibinfo {author} {\bibnamefont {{D. V. Lang}}},\
  }\bibfield  {title} {\bibinfo {title} {{Deep‐level transient spectroscopy:
  A new method to characterize traps in semiconductors}},\ }\href
  {https://doi.org/10.1063/1.1663719} {\bibfield  {journal} {\bibinfo
  {journal} {{J. Appl. Phys.}}\ }\textbf {\bibinfo {volume} {45}},\ \bibinfo
  {pages} {3023} (\bibinfo {year} {1974})}\BibitemShut {NoStop}%
\bibitem [{\citenamefont {{S. M. Myers, P. J. Cooper, and W. R.
  Wampler}}(2008)}]{myers2008}%
  \BibitemOpen
  \bibfield  {author} {\bibinfo {author} {\bibnamefont {{S. M. Myers, P. J.
  Cooper, and W. R. Wampler}}},\ }\bibfield  {title} {\bibinfo {title} {{Model
  of defect reactions and the influence of clustering in
  pulse-neutron-irradiated Si}},\ }\href {https://doi.org/10.1063/1.2963697}
  {\bibfield  {journal} {\bibinfo  {journal} {{J. Appl. Phys.}}\ }\textbf
  {\bibinfo {volume} {104}},\ \bibinfo {pages} {044507} (\bibinfo {year}
  {2008})}\BibitemShut {NoStop}%
\bibitem [{\citenamefont {{W. R. Wampler and S. M.
  Myers}}(2015)}]{wampler2015}%
  \BibitemOpen
  \bibfield  {author} {\bibinfo {author} {\bibnamefont {{W. R. Wampler and S.
  M. Myers}}},\ }\bibfield  {title} {\bibinfo {title} {{Model for transport and
  reaction of defects and carriers within displacement cascades in gallium
  arsenide}},\ }\href {https://doi.org/10.1063/1.4906104} {\bibfield  {journal}
  {\bibinfo  {journal} {J. Appl. Phys.}\ }\textbf {\bibinfo {volume} {117}},\
  \bibinfo {pages} {045707} (\bibinfo {year} {2015})}\BibitemShut {NoStop}%
\bibitem [{\citenamefont {{L. Musson, G. L. Hennigan, X. Gao, R. Humphreys, M.
  Negoita, and A. Huang, Sandia National Laboratory}}()}]{charon}%
  \BibitemOpen
  \bibfield  {author} {\bibinfo {author} {\bibnamefont {{L. Musson, G. L.
  Hennigan, X. Gao, R. Humphreys, M. Negoita, and A. Huang, Sandia National
  Laboratory}}},\ }\href@noop {} {\bibinfo {title} {{Charon User Manual V.2.1
  (Rev.01)}}},\ \bibinfo {note} {{SAND2022-7653, 2022}}\BibitemShut {NoStop}%
\bibitem [{\citenamefont {{G. M. Loubriel, F. J. Zutavern, A. G. Baca, H. P.
  Hjalmarson, T. A. Plut, W. D. Helgeson, M. W. O'Malley, M. H. Ruebush, D. J.
  and Brown}}(1997)}]{loubriel1997}%
  \BibitemOpen
  \bibfield  {author} {\bibinfo {author} {\bibnamefont {{G. M. Loubriel, F. J.
  Zutavern, A. G. Baca, H. P. Hjalmarson, T. A. Plut, W. D. Helgeson, M. W.
  O'Malley, M. H. Ruebush, D. J. and Brown}}},\ }\bibfield  {title} {\bibinfo
  {title} {{Photoconductive semiconductor switches}},\ }\href
  {https://doi.org/10.1109/27.602482} {\bibfield  {journal} {\bibinfo
  {journal} {{IEEE Trans. Plasma Sci.}}\ }\textbf {\bibinfo {volume} {25}},\
  \bibinfo {pages} {124} (\bibinfo {year} {1997})}\BibitemShut {NoStop}%
\bibitem [{\citenamefont {{K. Kambour, S. Kang, C. W. Myles, and H. P.
  Hjalmarson}}(2000)}]{kambour2000}%
  \BibitemOpen
  \bibfield  {author} {\bibinfo {author} {\bibnamefont {{K. Kambour, S. Kang,
  C. W. Myles, and H. P. Hjalmarson}}},\ }\bibfield  {title} {\bibinfo {title}
  {{Steady-state properties of lock-on current filaments in GaAs}},\ }\href
  {https://doi.org/10.1109/27.901221} {\bibfield  {journal} {\bibinfo
  {journal} {{IEEE Trans. Plasma Sci.}}\ }\textbf {\bibinfo {volume} {28}},\
  \bibinfo {pages} {1497} (\bibinfo {year} {2000})}\BibitemShut {NoStop}%
\bibitem [{\citenamefont {{V. Meyers, L. Voss, J. D. Flicker, L. G. Rodriguez,
  H. P. Hjalmarson, J. Lehr, N. Gonzalez, G. Pickrell, S. Ghandiparsi, and R.
  Kaplar}}(2025)}]{meyers2025}%
  \BibitemOpen
  \bibfield  {author} {\bibinfo {author} {\bibnamefont {{V. Meyers, L. Voss, J.
  D. Flicker, L. G. Rodriguez, H. P. Hjalmarson, J. Lehr, N. Gonzalez, G.
  Pickrell, S. Ghandiparsi, and R. Kaplar}}},\ }\bibfield  {title} {\bibinfo
  {title} {{Photoconductive Semiconductor Switches: Materials, Physics, and
  Applications}},\ }\href {https://doi.org/10.3390/app15020645} {\bibfield
  {journal} {\bibinfo  {journal} {{Appl. Sci.}}\ }\textbf {\bibinfo {volume}
  {15}},\ \bibinfo {pages} {645} (\bibinfo {year} {2025})}\BibitemShut
  {NoStop}%
\bibitem [{\citenamefont {{H. P. Hjalmarson, R. L. Pease, S. C. Witczak, M. R.
  Shaneyfelt, J. R. Schwank, A. H. Edwards, C. E. Hembree, and T. R.
  Mattsson}}(2003)}]{hjalmarson2003}%
  \BibitemOpen
  \bibfield  {author} {\bibinfo {author} {\bibnamefont {{H. P. Hjalmarson, R.
  L. Pease, S. C. Witczak, M. R. Shaneyfelt, J. R. Schwank, A. H. Edwards, C.
  E. Hembree, and T. R. Mattsson}}},\ }\bibfield  {title} {\bibinfo {title}
  {{Mechanisms for radiation dose-rate sensitivity of bipolar transistors}},\
  }\href {https://doi.org/10.1109/TNS.2003.821803} {\bibfield  {journal}
  {\bibinfo  {journal} {{IEEE Trans. Nuc. Sci.}}\ }\textbf {\bibinfo {volume}
  {50}},\ \bibinfo {pages} {1901} (\bibinfo {year} {2003})}\BibitemShut
  {NoStop}%
\bibitem [{\citenamefont {{R. L. Pease, P. C. Adell, B. G. Rax, X. J. Chen, H.
  J. Barnaby, K. E. Holbert, and H. P. Hjalmarson}}(2008)}]{pease2008}%
  \BibitemOpen
  \bibfield  {author} {\bibinfo {author} {\bibnamefont {{R. L. Pease, P. C.
  Adell, B. G. Rax, X. J. Chen, H. J. Barnaby, K. E. Holbert, and H. P.
  Hjalmarson}}},\ }\bibfield  {title} {\bibinfo {title} {{The Effects of
  Hydrogen on the Enhanced Low Dose Rate Sensitivity (ELDRS) of Bipolar Linear
  Circuits}},\ }\href {https://doi.org/10.1109/TNS.2008.2006478} {\bibfield
  {journal} {\bibinfo  {journal} {{IEEE Trans. Nuc. Sci.}}\ }\textbf {\bibinfo
  {volume} {55}},\ \bibinfo {pages} {3169} (\bibinfo {year}
  {2008})}\BibitemShut {NoStop}%
\bibitem [{\citenamefont {{H. P. Hjalmarson, R. L. Pease, and R. A. B.
  Devine}}(2008)}]{hjalmarson2008}%
  \BibitemOpen
  \bibfield  {author} {\bibinfo {author} {\bibnamefont {{H. P. Hjalmarson, R.
  L. Pease, and R. A. B. Devine}}},\ }\bibfield  {title} {\bibinfo {title}
  {{Calculations of Radiation Dose-Rate Sensitivity of Bipolar Transistors}},\
  }\href {https://doi.org/10.1109/TNS.2008.2007487} {\bibfield  {journal}
  {\bibinfo  {journal} {{IEEE Trans. Nuc. Sci.}}\ }\textbf {\bibinfo {volume}
  {55}},\ \bibinfo {pages} {3009} (\bibinfo {year} {2008})}\BibitemShut
  {NoStop}%
\bibitem [{\citenamefont {{B.D. Tierney, H. P. Hjalmarson, R. B. Jacobs-Gedrim,
  S. Agarwal, C. D. James, and M. J. Marinella}}(2018)}]{tierney2018}%
  \BibitemOpen
  \bibfield  {author} {\bibinfo {author} {\bibnamefont {{B.D. Tierney, H. P.
  Hjalmarson, R. B. Jacobs-Gedrim, S. Agarwal, C. D. James, and M. J.
  Marinella}}},\ }\bibfield  {title} {\bibinfo {title} {{Unified computational
  model of transport in metal-insulating oxide-metal systems}},\ }\bibfield
  {journal} {\bibinfo  {journal} {{Appl. Phys. A}}\ }\textbf {\bibinfo {volume}
  {124}},\ \href {https://doi.org/10.1007/s00339-018-1632-3}
  {10.1007/s00339-018-1632-3} (\bibinfo {year} {2018})\BibitemShut {NoStop}%
\bibitem [{\citenamefont {{W. Shockley}}(1949)}]{shockley1949}%
  \BibitemOpen
  \bibfield  {author} {\bibinfo {author} {\bibnamefont {{W. Shockley}}},\
  }\bibfield  {title} {\bibinfo {title} {{The theory of p-n junctions in
  semiconductors and p-n junction transistors}},\ }\href
  {https://doi.org/10.1002/j.1538-7305.1949.tb03645.x} {\bibfield  {journal}
  {\bibinfo  {journal} {The Bell Sys. Tech. J.}\ }\textbf {\bibinfo {volume}
  {28}},\ \bibinfo {pages} {435} (\bibinfo {year} {1949})}\BibitemShut
  {NoStop}%
\bibitem [{\citenamefont {Roosbroeck}(1950)}]{roosbroeck1950}%
  \BibitemOpen
  \bibfield  {author} {\bibinfo {author} {\bibfnamefont {W.~V.}\ \bibnamefont
  {Roosbroeck}},\ }\bibfield  {title} {\bibinfo {title} {{Theory of the flow of
  electrons and holes in germanium and other semiconductors}},\ }\href
  {https://doi.org/10.1002/j.1538-7305.1950.tb03653.x} {\bibfield  {journal}
  {\bibinfo  {journal} {Bell Sys. Tech. J.}\ }\textbf {\bibinfo {volume}
  {29}},\ \bibinfo {pages} {560} (\bibinfo {year} {1950})}\BibitemShut
  {NoStop}%
\bibitem [{\citenamefont {{R. M. Fleming, D. V. Lang, C. H. Seager, E.
  Bielejec, G. A. Patrizi, and J. M. Campbell}}(2010)}]{fleming2010}%
  \BibitemOpen
  \bibfield  {author} {\bibinfo {author} {\bibnamefont {{R. M. Fleming, D. V.
  Lang, C. H. Seager, E. Bielejec, G. A. Patrizi, and J. M. Campbell}}},\
  }\bibfield  {title} {\bibinfo {title} {{Continuous distribution of defect
  states and band gap narrowing in neutron irradiated GaAs}},\ }\href
  {https://doi.org/10.1063/1.3448118} {\bibfield  {journal} {\bibinfo
  {journal} {J. Appl. Phys.}\ }\textbf {\bibinfo {volume} {107}},\ \bibinfo
  {pages} {123710} (\bibinfo {year} {2010})}\BibitemShut {NoStop}%
\bibitem [{\citenamefont {{J. Dabrowski and M.
  Scheffler}}(1988)}]{dabrowski1988}%
  \BibitemOpen
  \bibfield  {author} {\bibinfo {author} {\bibnamefont {{J. Dabrowski and M.
  Scheffler}}},\ }\bibfield  {title} {\bibinfo {title} {{Theoretical Evidence
  for an Optically Inducible Structural Transition of the Isolated As Antisite
  in GaAs: Identification and Explanation of $\mathrm{EL}2$?}},\ }\href
  {https://doi.org/10.1103/PhysRevLett.60.2183} {\bibfield  {journal} {\bibinfo
   {journal} {Phys. Rev. Lett.}\ }\textbf {\bibinfo {volume} {60}},\ \bibinfo
  {pages} {2183} (\bibinfo {year} {1988})}\BibitemShut {NoStop}%
\bibitem [{\citenamefont {{D. J. Chadi and K. J. Chang}}(1988)}]{chadi1988}%
  \BibitemOpen
  \bibfield  {author} {\bibinfo {author} {\bibnamefont {{D. J. Chadi and K. J.
  Chang}}},\ }\bibfield  {title} {\bibinfo {title} {{Metastability of the
  Isolated Arsenic-Antisite Defect in GaAs}},\ }\href
  {https://doi.org/10.1103/PhysRevLett.60.2187} {\bibfield  {journal} {\bibinfo
   {journal} {Phys. Rev. Lett.}\ }\textbf {\bibinfo {volume} {60}},\ \bibinfo
  {pages} {2187} (\bibinfo {year} {1988})}\BibitemShut {NoStop}%
\bibitem [{\citenamefont {{P. A. Schultz}}(2006)}]{schultz2006}%
  \BibitemOpen
  \bibfield  {author} {\bibinfo {author} {\bibnamefont {{P. A. Schultz}}},\
  }\bibfield  {title} {\bibinfo {title} {{Theory of Defect Levels and the
  ``Band Gap Problem'' in Silicon}},\ }\href
  {https://doi.org/10.1103/PhysRevLett.96.246401} {\bibfield  {journal}
  {\bibinfo  {journal} {Phys. Rev. Lett.}\ }\textbf {\bibinfo {volume} {96}},\
  \bibinfo {pages} {246401} (\bibinfo {year} {2006})}\BibitemShut {NoStop}%
\bibitem [{\citenamefont {{A. H. Edwards, P. A. Schultz, and R. M.
  Dobzynski}}(2022)}]{edwards2022}%
  \BibitemOpen
  \bibfield  {author} {\bibinfo {author} {\bibnamefont {{A. H. Edwards, P. A.
  Schultz, and R. M. Dobzynski}}},\ }\bibfield  {title} {\bibinfo {title}
  {{Electronic structure of intrinsic defects in $c$-gallium nitride: Density
  functional theory study without the jellium approximation}},\ }\href
  {https://doi.org/10.1103/PhysRevB.105.235110} {\bibfield  {journal} {\bibinfo
   {journal} {Phys. Rev. B}\ }\textbf {\bibinfo {volume} {105}},\ \bibinfo
  {pages} {235110} (\bibinfo {year} {2022})}\BibitemShut {NoStop}%
\bibitem [{\citenamefont {{D. Pons, A. Mircea, and J.
  Bourgoin}}(1980)}]{pons1980}%
  \BibitemOpen
  \bibfield  {author} {\bibinfo {author} {\bibnamefont {{D. Pons, A. Mircea,
  and J. Bourgoin}}},\ }\bibfield  {title} {\bibinfo {title} {{An annealing
  study of electron irradiation‐induced defects in GaAs}},\ }\href
  {https://doi.org/10.1063/1.328235} {\bibfield  {journal} {\bibinfo  {journal}
  {Journal of Applied Physics}\ }\textbf {\bibinfo {volume} {51}},\ \bibinfo
  {pages} {4150} (\bibinfo {year} {1980})}\BibitemShut {NoStop}%
\bibitem [{\citenamefont {{P. Pichler}}(2004)}]{pichler}%
  \BibitemOpen
  \bibfield  {author} {\bibinfo {author} {\bibnamefont {{P. Pichler}}},\
  }\href@noop {} {\emph {\bibinfo {title} {{Intrinsic Point Defects,
  Impurities, and Their Diffusion in Silicon}}}}\ (\bibinfo  {publisher}
  {Springer, New York},\ \bibinfo {year} {2004})\BibitemShut {NoStop}%
\bibitem [{\citenamefont {{J. L. Benton and L. C.
  Kimerling}}(1982)}]{Benton_1982}%
  \BibitemOpen
  \bibfield  {author} {\bibinfo {author} {\bibnamefont {{J. L. Benton and L. C.
  Kimerling}}},\ }\bibfield  {title} {\bibinfo {title} {{Capacitance Transient
  Spectroscopy of Trace Contamination in Silicon}},\ }\href
  {https://doi.org/10.1149/1.2124387} {\bibfield  {journal} {\bibinfo
  {journal} {J. Electrochem. Soc.}\ }\textbf {\bibinfo {volume} {129}},\
  \bibinfo {pages} {2098} (\bibinfo {year} {1982})}\BibitemShut {NoStop}%
\bibitem [{\citenamefont {{J.C Bourgoin and J.W
  Corbett}}(1972)}]{bourgoin-corbett}%
  \BibitemOpen
  \bibfield  {author} {\bibinfo {author} {\bibnamefont {{J.C Bourgoin and J.W
  Corbett}}},\ }\bibfield  {title} {\bibinfo {title} {{A new mechanism for
  interstistitial migration}},\ }\href
  {https://doi.org/https://doi.org/10.1016/0375-9601(72)90523-3} {\bibfield
  {journal} {\bibinfo  {journal} {Phys. Lett. A}\ }\textbf {\bibinfo {volume}
  {38}},\ \bibinfo {pages} {135} (\bibinfo {year} {1972})}\BibitemShut
  {NoStop}%
\bibitem [{\citenamefont {{I. Vurgaftman, J. R. Meyer, and L. R.
  Ram-Mohan}}(2001)}]{vurgaftman2001}%
  \BibitemOpen
  \bibfield  {author} {\bibinfo {author} {\bibnamefont {{I. Vurgaftman, J. R.
  Meyer, and L. R. Ram-Mohan}}},\ }\bibfield  {title} {\bibinfo {title} {{Band
  parameters for III–V compound semiconductors and their alloys}},\ }\href
  {https://doi.org/https://doi.org/10.1063/1.1368156} {\bibfield  {journal}
  {\bibinfo  {journal} {J. Appl. Phys.}\ }\textbf {\bibinfo {volume} {89}},\
  \bibinfo {pages} {5815} (\bibinfo {year} {2001})}\BibitemShut {NoStop}%
\bibitem [{\citenamefont {{J. S. Blakemore}}(1982)}]{blakemore1982}%
  \BibitemOpen
  \bibfield  {author} {\bibinfo {author} {\bibnamefont {{J. S. Blakemore}}},\
  }\bibfield  {title} {\bibinfo {title} {{Semiconducting and other major
  properties of gallium arsenide}},\ }\href {https://doi.org/10.1063/1.331665}
  {\bibfield  {journal} {\bibinfo  {journal} {J. Appl. Phys.}\ }\textbf
  {\bibinfo {volume} {53}},\ \bibinfo {pages} {R123} (\bibinfo {year}
  {1982})}\BibitemShut {NoStop}%
\bibitem [{\citenamefont {Neuberger}(1971)}]{neuberger}%
  \BibitemOpen
  \bibfield  {author} {\bibinfo {author} {\bibfnamefont {N.}~\bibnamefont
  {Neuberger}},\ }\href@noop {} {\emph {\bibinfo {title} {Handbook of
  Electronic Materials}}},\ Vol.~\bibinfo {volume} {5}\ (\bibinfo  {publisher}
  {Plenum, New York},\ \bibinfo {year} {1971})\BibitemShut {NoStop}%
\bibitem [{\citenamefont {{P. L\"oper, R. M\"uller, D. Hiller, T. Barthel, E.
  Malguth, S. Janz, J. C. Goldschmidt, M. Hermle, and M.
  Zacharias}}(2011)}]{loper2011}%
  \BibitemOpen
  \bibfield  {author} {\bibinfo {author} {\bibnamefont {{P. L\"oper, R.
  M\"uller, D. Hiller, T. Barthel, E. Malguth, S. Janz, J. C. Goldschmidt, M.
  Hermle, and M. Zacharias}}},\ }\bibfield  {title} {\bibinfo {title}
  {{Quasi-Fermi-level splitting in ideal silicon nanocrystal superlattices}},\
  }\href {https://doi.org/10.1103/PhysRevB.84.195317} {\bibfield  {journal}
  {\bibinfo  {journal} {Phys. Rev. B}\ }\textbf {\bibinfo {volume} {84}},\
  \bibinfo {pages} {195317} (\bibinfo {year} {2011})}\BibitemShut {NoStop}%
\bibitem [{\citenamefont {{N. Jain, R. Saxena, S. Vaidya, W. Huang, A. Welford,
  C. R. McNeill, and D. Kabra}}(2022)}]{jain2022}%
  \BibitemOpen
  \bibfield  {author} {\bibinfo {author} {\bibnamefont {{N. Jain, R. Saxena, S.
  Vaidya, W. Huang, A. Welford, C. R. McNeill, and D. Kabra}}},\ }\bibfield
  {title} {\bibinfo {title} {{Light induced quasi-Fermi level splitting in
  molecular semiconductor alloys}},\ }\href
  {https://doi.org/10.1039/D2MA00131D} {\bibfield  {journal} {\bibinfo
  {journal} {Mater. Adv.}\ }\textbf {\bibinfo {volume} {3}},\ \bibinfo {pages}
  {5344} (\bibinfo {year} {2022})}\BibitemShut {NoStop}%
\bibitem [{\citenamefont {{P. Reddy, F. Kaess, J. Tweedie, R. Kirste, S. Mita,
  R. Collazo, and Z. Sitar}}(2017)}]{Reddy2017}%
  \BibitemOpen
  \bibfield  {author} {\bibinfo {author} {\bibnamefont {{P. Reddy, F. Kaess, J.
  Tweedie, R. Kirste, S. Mita, R. Collazo, and Z. Sitar}}},\ }\bibfield
  {title} {\bibinfo {title} {{Defect quasi Fermi level control-based C$_N$
  reduction in GaN: Evidence for the role of minority carriers}},\ }\href
  {https://doi.org/10.1063/1.5000720} {\bibfield  {journal} {\bibinfo
  {journal} {Appl. Phys. Lett.}\ }\textbf {\bibinfo {volume} {111}},\ \bibinfo
  {pages} {152101} (\bibinfo {year} {2017})}\BibitemShut {NoStop}%
\bibitem [{\citenamefont {{D. B. Riley, O. J. Sandberg, N. M. Wilson, W. Li, S.
  Zeiske, N. Zarrabi, P. Meredith, R. \"Osterbacka, Ronald and A.
  Armin}}(2021)}]{Riley2021}%
  \BibitemOpen
  \bibfield  {author} {\bibinfo {author} {\bibnamefont {{D. B. Riley, O. J.
  Sandberg, N. M. Wilson, W. Li, S. Zeiske, N. Zarrabi, P. Meredith, R.
  \"Osterbacka, Ronald and A. Armin}}},\ }\bibfield  {title} {\bibinfo {title}
  {{Direct Quantification of Quasi-Fermi-Level Splitting in Organic
  Semiconductor Devices}},\ }\href
  {https://doi.org/10.1103/PhysRevApplied.15.064035} {\bibfield  {journal}
  {\bibinfo  {journal} {Phys. Rev. Appl.}\ }\textbf {\bibinfo {volume} {15}},\
  \bibinfo {pages} {064035} (\bibinfo {year} {2021})}\BibitemShut {NoStop}%
\bibitem [{\citenamefont {{Lombez, Laurent and Paire, Myriam and Ory, Daniel
  and Delamarre, Amaury and Rodière, Jean and Rale, Pierre and El-Hajje,
  Gilbert and Guillemoles, Jean-François}}(2014)}]{Lombez2014}%
  \BibitemOpen
  \bibfield  {author} {\bibinfo {author} {\bibnamefont {{Lombez, Laurent and
  Paire, Myriam and Ory, Daniel and Delamarre, Amaury and Rodière, Jean and
  Rale, Pierre and El-Hajje, Gilbert and Guillemoles, Jean-François}}},\
  }\bibfield  {title} {\bibinfo {title} {{Direct imaging of quasi Fermi level
  splitting in photovoltaic absorbers}},\ }in\ \href
  {https://doi.org/10.1109/PVSC.2014.6925016} {\emph {\bibinfo {booktitle}
  {2014 IEEE 40th Photovoltaic Specialist Conference (PVSC)}}}\ (\bibinfo
  {year} {2014})\ p.\ \bibinfo {pages} {0695}\BibitemShut {NoStop}%
\bibitem [{\citenamefont {{L. Q. Phuong, S. M. Hosseini, O. J. Sandberg, Y.
  Zou, H. Y. Woo, D. Neher, and S. Shoaee}}(2021)}]{phuong2021}%
  \BibitemOpen
  \bibfield  {author} {\bibinfo {author} {\bibnamefont {{L. Q. Phuong, S. M.
  Hosseini, O. J. Sandberg, Y. Zou, H. Y. Woo, D. Neher, and S. Shoaee}}},\
  }\bibfield  {title} {\bibinfo {title} {{Quantifying Quasi-Fermi Level
  Splitting and Open-Circuit Voltage Losses in Highly Efficient Nonfullerene
  Organic Solar Cells}},\ }\href
  {https://doi.org/https://doi.org/10.1002/solr.202000649} {\bibfield
  {journal} {\bibinfo  {journal} {Solar RRL}\ }\textbf {\bibinfo {volume}
  {5}},\ \bibinfo {pages} {2000649} (\bibinfo {year} {2021})}\BibitemShut
  {NoStop}%
\bibitem [{\citenamefont {{J. Warby, S. Shah, J. Thiesbrummel, E.
  Gutierrez-Partida, H. Lai, B. Alebachew, M. Grischek, F. Yang, F. Lang, S.
  Albrecht, F. Fu, D. Neher, and M. Stolterfoht}}(2023)}]{warby2023}%
  \BibitemOpen
  \bibfield  {author} {\bibinfo {author} {\bibnamefont {{J. Warby, S. Shah, J.
  Thiesbrummel, E. Gutierrez-Partida, H. Lai, B. Alebachew, M. Grischek, F.
  Yang, F. Lang, S. Albrecht, F. Fu, D. Neher, and M. Stolterfoht}}},\
  }\bibfield  {title} {\bibinfo {title} {{Mismatch of Quasi–Fermi Level
  Splitting and Voc in Perovskite Solar Cells}},\ }\href
  {https://doi.org/https://doi.org/10.1002/aenm.202303135} {\bibfield
  {journal} {\bibinfo  {journal} {Adv. Energy Mater.}\ }\textbf {\bibinfo
  {volume} {13}},\ \bibinfo {pages} {2303135} (\bibinfo {year}
  {2023})}\BibitemShut {NoStop}%
\bibitem [{\citenamefont {{Z. -Y. Yan, K. -H. Xue, Z. Hou, Y. Shen, H. Tian, Y.
  Yang, and T. -L. Ren}}(2022)}]{yan2022}%
  \BibitemOpen
  \bibfield  {author} {\bibinfo {author} {\bibnamefont {{Z. -Y. Yan, K. -H.
  Xue, Z. Hou, Y. Shen, H. Tian, Y. Yang, and T. -L. Ren}}},\ }\bibfield
  {title} {\bibinfo {title} {{Quasi-Fermi-Level Phase Space and its
  Applications in Ambipolar Two-Dimensional Field-Effect Transistors}},\ }\href
  {https://doi.org/10.1103/PhysRevApplied.17.054027} {\bibfield  {journal}
  {\bibinfo  {journal} {Phys. Rev. Appl.}\ }\textbf {\bibinfo {volume} {17}},\
  \bibinfo {pages} {054027} (\bibinfo {year} {2022})}\BibitemShut {NoStop}%
\bibitem [{\citenamefont {{K. Thommen}}(1970)}]{thommen1970}%
  \BibitemOpen
  \bibfield  {author} {\bibinfo {author} {\bibnamefont {{K. Thommen}}},\
  }\bibfield  {title} {\bibinfo {title} {{Recovery of low temperature electron
  irradiation-induced damage in n-type GaAs}},\ }\href
  {https://doi.org/10.1080/00337577008243053} {\bibfield  {journal} {\bibinfo
  {journal} {Radiat. Eff.}\ }\textbf {\bibinfo {volume} {2}},\ \bibinfo {pages}
  {201} (\bibinfo {year} {1970})}\BibitemShut {NoStop}%
\end{thebibliography}%

\appendix
\section{Units}

\begin{table}[h]
\renewcommand{\arraystretch}{1.3}
\caption{\label{tab:variables} Parameter, variable, and units for quantities used in this work.}
\centering
\begin{tabularx}{\columnwidth}{lcc}
\hline
\bfseries Parameter &
\bfseries Variable&
{\bfseries Unit}\\
\hline\hline
species & $c_{i}$ & cm$^{-3}$  \\
time & $t$ & s  \\
current density & $J_{i}$& A/cm$^2$  \\
position & $\textbf{r}$ & cm \\
stoichiometric coefficient & $\gamma_{ij}$ & unitless  \\
reaction j & $r_{j}$ & unitless  \\
mobility & $\mu_{i}$ & cm$^2/$V$\cdot$ s  \\
charge & $q$ & C  \\
electrochemical potential & $\Phi_i(\textbf{r})$ & eV \\
electrostatic potential & $\phi(\textbf{r})$ & V  \\
chemical potential  & $\nu_i(\textbf{r})$ & eV  \\
charge density & $\rho(\textbf{r})$ & C/cm$^3$  \\
dielectric coefficient & $\varepsilon$ & unitless   \\
Fermi Level & $E_F$ & eV  \\
electron quasi-Fermi Level & $E_F^e$ & eV  \\
hole quasi-Fermi Level & $E_F^h$ & eV  \\
Conduction band density of states & $N_C$ & cm$^{-3}$ \\
Valence band density of states & $N_V$ & cm$^{-3}$  \\
electron effective mass & $m_e^*$ & kg \\
hole effective mass & $m_h^*$ & kg \\
Planck constant & $\hbar$ & eV$\cdot$ s\\
\hline
\end{tabularx}
\end{table}

\section{Other Features}
In addition to the features presented here, the REOS suite can implement Auger and Shockley-Reed-Hall (SRH) recombination (not used in this work).


\end{document}